\documentclass[twocolumn]{aastex61}

\turnoffedit

\usepackage{amsmath}
\usepackage{amssymb}
\usepackage{verbatim}
\usepackage{subfigure}

\newcommand{\E}[1]{\ensuremath{\times 10^{#1}}}

\newcommand{\de}{\ensuremath{\mathrm{d}}}
\newcommand{\dee}{\ensuremath{\ \mathrm{d}}}
\newcommand{\sersic}{S\'{e}rsic}
\newcommand{\Reff}{\ensuremath{R_\mathrm{e}}}
\newcommand{\todo}[1]{\authorcomment1{#1}}

\shorttitle{The Inner Dark Matter Density Slope of NGC~1407}
\shortauthors{Wasserman et al.}

\begin{document}

\title{The SLUGGS Survey: The Inner Dark Matter Density Slope of the Massive Elliptical Galaxy NGC~1407}

\correspondingauthor{Asher Wasserman}
\email{adwasser@ucsc.edu}

\author{Asher Wasserman}
\affil{Department of Astronomy \& Astrophysics, University of California-Santa Cruz, Santa Cruz, CA 95064, USA}
\author{Aaron J. Romanowsky}
\affil{Department of Physics and Astronomy, San Jos\'{e} State University, One Washington Square, San Jose, CA 95192, USA}
\affil{University of California Observatories, 1156 High Street, Santa Cruz, CA 95064, USA}
\author{Jean Brodie}
\affil{Department of Astronomy \& Astrophysics, University of California-Santa Cruz, Santa Cruz, CA 95064, USA}
\affil{University of California Observatories, 1156 High Street, Santa Cruz, CA 95064, USA}
\author{Pieter van Dokkum}
\affil{Astronomy Department, Yale University, New Haven, CT 06511, USA}
\author{Charlie Conroy}
\affil{Department of Astronomy, Harvard University, Cambridge, MA 02138, USA}
\author{Alexa Villaume}
\affil{Department of Astronomy \& Astrophysics, University of California-Santa Cruz, Santa Cruz, CA 95064, USA}
\author{Duncan A. Forbes}
\affil{Centre for Astrophysics \& Supercomputing, Swinburne University, Hawthorn VIC 3122, Australia}
\author{Jay Strader}
\affil{Center for Data Intensive and Time Domain Astronomy, Department of Physics and Astronomy, Michigan State University, East Lansing, MI 48824, USA}
\author{Adebusola Alabi}
\affil{University of California Observatories, 1156 High Street, Santa Cruz, CA 95064, USA}
\author{Sabine Bellstedt}
\affil{Centre for Astrophysics \& Supercomputing, Swinburne University, Hawthorn VIC 3122, Australia}

\begin{abstract}
  We investigate the dark matter density profile of the massive elliptical galaxy, NGC~1407, by constructing spherically symmetric Jeans models of its field star and globular cluster systems.  Two major challenges in such models are the degeneracy between the stellar mass and the dark matter halo profiles, and the degeneracy between the orbital anisotropy of the tracer population and the total mass causing the observed motions.  We address the first issue by using new measurements of the mass-to-light ratio profile from stellar population constraints that include a radially varying initial mass function.  To deal with the mass--anisotropy degeneracy, we make use of multiple kinematic tracers, including two subpopulations of globular clusters in addition to the galaxy's field stars.  We create a hierarchical Bayesian model that addresses several often neglected systematic uncertainties, such as the statistical weight given to various datasets and the adopted distance.  After sampling the posterior probability distribution with a Markov Chain Monte Carlo method, we find evidence for a central cusp with a log-slope of $\gamma = 1.0^{+0.3}_{-0.4}$.  This is lower than expected for dark matter halos that have undergone adiabatic contraction, supporting inferences from gravitational lensing that some process has suppressed the steepening of halos in massive galaxies.  We also confirm radially-biased orbits for the metal-rich globular clusters and tangentially-biased orbits for the metal-poor globular clusters, which remains a puzzling finding for an accretion-dominated halo.
\end{abstract}


\section{Introduction}
\label{sec:intro}

The concordance cosmological model of dark energy plus cold dark matter ($\Lambda$CDM) has had numerous successes in describing the large scale structure of the universe.  The story on the scale of galaxy formation has been more complicated, with discrepancies in the number of satellite galaxies expected around the Milky Way \citep{klypin+1999, moore+1999}, the masses of the Milky Way satellites that are observed \citep{boylan-kolchin+2011, bullock&boylan-kolchin2017}, and the inner slope of the dark matter density profile of galaxies \citep{flores&primack1994}.  It is this last point that we focus on here.

\cite*{navarro+1997} introduced a double power law model (hereafter the NFW model) of the halo density profile with $\rho \propto r^{-1}$ in the inner regions and $\rho \propto r^{-3}$ in the outer regions, which they found to describe well the form of halos from N-body simulations. This model can be generalized to include a variable inner slope, and is often parameterized as
\begin{equation}
  \label{eqn:nfw}
  \rho(r) = \rho_s \left(\frac{r}{r_s}\right) ^ {-\gamma} \left(1 + \frac{r}{r_s}\right) ^ {\gamma - 3} \ ,
\end{equation}
where $r_s$ is the scale radius which determines where the change in density slope occurs.

For $\gamma = 1$, this corresponds to the original NFW profile.  While this ``universal'' profile provided a good match to their DM-only simulations, deviations from this profile have been observed in various mass regimes.  For instance, dwarf galaxies have often been found to have shallower inner density slopes (\citealt{simon+2003, spekkens+2005, walker&penarrubia2011, oh+2011a}, though see \citealt{adams+2014, pineda+2017}).  On the opposite end of the mass spectrum, \cite{newman+2013b} used both gravitational lensing and stellar dynamics to measure $\langle \gamma \rangle \sim 0.5$ for a sample of massive galaxy clusters.

If DM halos start with an NFW-like steep inner profile, than some physical mechanism for transferring energy to DM in the inner regions is necessary to create the shallower DM profiles observed for some galaxies.  Self-interacting or fuzzy DM scenarios have been proposed to solve this issue (e.g, \citealt{rocha+2013, robles+2015, diCintio+2017}).  However, baryonic effects may also explain DM cores, either from bursty star formation at the low mass end \citep{navarro+1996, mashchenko+2008, pontzen&governato2012} or from dynamical friction during gas-poor mergers at the high mass end \citep{el-zant+2004}.  In addition, \cite{dekel+2003a} argued that merging satellites whose halos have DM cores would be disrupted outside of the central halo's core, leading to a stable DM core in the central galaxy.  Whatever processes are responsible for flattening the DM density profile must compete with the effects of adiabatic contraction \citep{blumenthal+1986}, whereby the infalling of gas during the process of galaxy formation causes a steepening of the halo density profile.

To disentangle these many effects on the halo, we need to observationally map out how the inner DM slope changes as a function of halo mass across a wide range of mass regimes.  While there are already many good constraints on this relation for dwarf galaxies and for clusters of galaxies, there remains a dearth of observational measurements of the inner DM slope for halos between the masses of $10^{12}$ and $10 ^ {13}$ $M_\odot$, which typically host massive early-type galaxies (ETGs).  These massive galaxies are particularly critical tests for the presence of new, non-CDM physics, as many of the baryonic effects on the halo are small compared to those for dwarf galaxies.

Mass inferences with dispersion-dominated dynamics suffer from a number of challenges.  For one, the total mass is degenerate with the distribution of the orbits of the kinematic tracers being modeled. A general strategy for dealing with this mass--anisotropy degeneracy is to simultaneously model multiple kinematic tracers with separate distributions of their orbits.

\cite{walker&penarrubia2011} applied this approach to the Fornax and Sculptor dwarf spheroidal (dSph) galaxies by splitting their resolved stellar kinematic data into chemo-dynamically distinct components, then making separate mass estimates using each subpopulation.  Previous studies of massive ETGs modeled multiple tracer populations such as globular clusters (GCs), planetary nebulae (PNe), and integrated-light stellar kinematics to alleviate the mass--anisotropy degeneracy \citep{schuberth+2010, agnello+2014, pota+2015b, zhu+2016, oldham+2016}.

These studies were able to provide good constraints on the total mass of DM halos, but inferring the detailed density distribution of halos requires a precise determination of the stellar mass distribution.  As \cite{pota+2015b} found, the degeneracy between the inferred stellar mass-to-light ratio ($\Upsilon_*$) and the inner DM density slope undermines attempts to draw robust conclusions about the slope of the DM halo.  Furthermore, in all of the studies cited above, $\Upsilon_*$ was assumed to be constant across all galactocentric radii (but see \citealt{li+2017, poci+2017, mitzkus+2017} for work that relaxes this assumption).  Given that many ETGs are found to have spatially varying stellar populations, the constant $\Upsilon_*$ assumption is an important systematic uncertainty in understanding the inner DM density distribution \citep{martinNavarro+2015, mcConnell+2016, vanDokkum+2017}.

Using data from the SAGES Legacy Unifying Globulars and GalaxieS (SLUGGS) survey\footnote{\url{http://sluggs.ucolick.org}} \citep{brodie+2014}, we model the dynamics of the massive elliptical galaxy NGC~1407.  SLUGGS is a survey of 25 nearby ETGs across a variety of masses, environments, and morphologies.  NGC~1407 has been studied by numerous authors (e.g., \citealt{romanowsky+2009, su+2014, pota+2015b}), but here we revisit the galaxy with state-of-the-art stellar population synthesis results, a new method for modeling the stellar mass distribution, and a more rigorous statistical treatment of the influence of multiple disparate datasets.  This paper is a pilot work for an expanded study of a larger subset of SLUGGS galaxies.

\todo{more discussion of previous dynamical modeling efforts for N1407}

In Section~\ref{sec:data}, we summarize the observational data.  In Section~\ref{sec:methods}, we describe the dynamical modeling and our method for combining distinct observational constraints.  In Section~\ref{sec:results} we present the results of applying our model to NGC~1407.  In Section~\ref{sec:discussion} we interpret these results in the context of other observations and theoretical predictions.  We summarize our findings in Section~\ref{sec:conclusions}, and we present our full posterior probability distributions in Appendix~\ref{sec:post}.

\section{Data}
\label{sec:data}

NGC~1407 is a bright ($M_K = -25.46$; \citealt{jarrett+2000}), X-ray luminous ($L_X = 8.6\E{40}$ erg s$^{-1}$ within 2 \Reff{}; \citealt{su&irwin2013}), massive elliptical galaxy at the center of its eponymous galaxy group. \cite{brough+2006a} argued on the basis of its high X-ray luminosity and low spiral fraction that the NGC~1407 group is dynamically mature.  The central galaxy is a slow rotator ($\lambda_{\Reff{}} = 0.09$; \citealt{bellstedt+2017a}).  We adopt a systemic velocity of 1779 km~s$^{-1}$ \citep{quintana+1994}.  The galaxy shows slight ellipticity (\citealt{rc3} reported a flattening of $\epsilon = 0.07$), and so we calculate the projected galactocentric radius as
\begin{align}
  R^2 = q \Delta x^2 + q^{-1} \Delta y^2 \ ,
\end{align}
where $\Delta x$ and $\Delta y$ are coordinate offsets along the major and minor axes respectively and $q$ is the axial ratio ($b/a$).  Here we have adopted a position angle of 58.4\degr \citep{spolaor+2008a}.

There are numerous conflicting redshift-independent distances for NGC~1407 in the literature.  \cite{cantiello+2005} used surface brightness fluctuation (SBF) measurements to obtain a value of $25.1 \pm 1.2$ Mpc, while \cite{forbes+2006} used the globular cluster luminosity function to obtain a value of $21.2 \pm 0.9$ Mpc.  Using a weighted average of both SBFs and fits to the Fundamental Plane, \cite{tully+2013} derived a distance of $28.2 \pm 3.4$ Mpc.  Using the \cite{planck2015} cosmological parameters and correcting the recession velocity to the Virgo infall frame, the galaxy has a luminosity distance of $24.2 \pm 1.7$ Mpc.  When including the distance to the galaxy as a free parameter, we use a Gaussian prior with a mean of 26 Mpc and a standard deviation of 2 Mpc.  We find an a posteriori distance of $21.0^{+1.5}_{-1.4}$ Mpc (see Sec.~\ref{sec:results}) corresponding to a distance scale of 0.102 kpc per arcsecond.  It is this distance that we adopt for any distance-dependent results that are not already marginalized over this parameter.

Here we summarize the kinematic, photometric, and stellar population data which we use for our models.

\subsection{Stellar density}
\label{sec:stdens}

We use the same surface brightness profile as \cite{pota+2013}, who combined Subaru/Suprime-Cam $g$ band and {\it HST}/ACS F435 imaging into a single $B$ band profile out to $440\arcsec$. Masking out the core at $R < 2\arcsec$, they fitted a single \sersic{} component (Eqn.~\ref{eq:sersic}).
\begin{equation}
  \label{eq:sersic}
  I(R) = I_0 \exp\left(-b_n \left(\frac{R}{R_e}\right)^{1/n}\right)
\end{equation}
Here, $I_0$ is the central surface density, \Reff{} is the effective radius, $n$ is the \sersic{} index, and $b_n$ is a function of $n$ chosen such that $2 L(R_e) = L_\mathrm{tot}$ \citep[see Eqn.~18 in][for an asymptotic expansion of $b_n$]{ciotti&bertin1999}.  \cite{pota+2013} found an effective radius of $\Reff = 100\arcsec \pm 3\arcsec$, a \sersic{} index of $n = 4.67 \pm 0.15$, and a central surface brightness of $I_0 = 1.55\E{11}$ L$_{\odot, B}$ kpc$^{-2}$ (adopting a solar absolute magnitude of $M_{\odot, B} = 5.48$).

\begin{figure}
  \centering
  \includegraphics[width=\linewidth]{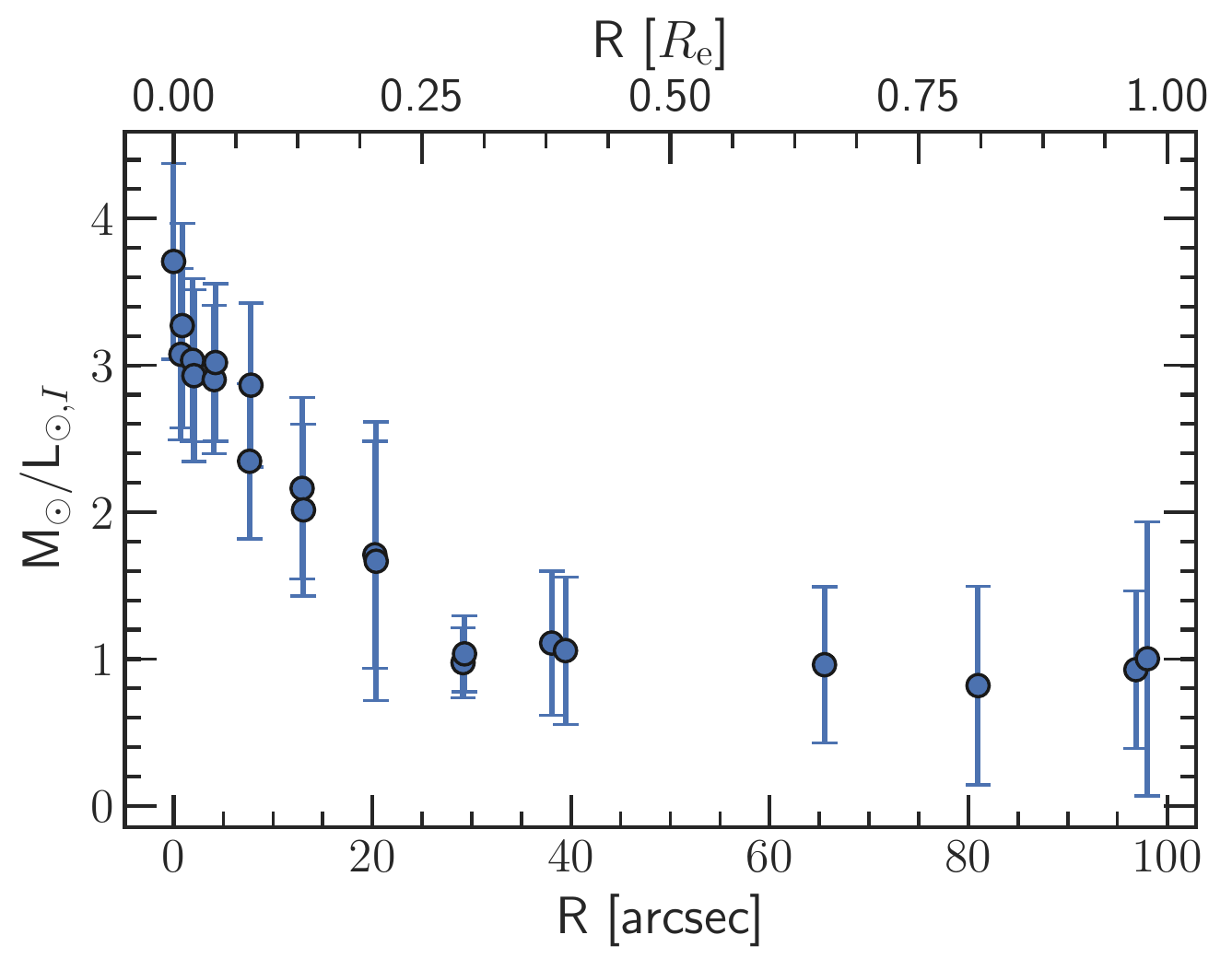}
  \caption{Stellar mass-to-light profile of NGC~1407 from \cite{vanDokkum+2017}.}
  \label{fig:ml_i}
\end{figure}

To derive a stellar mass surface density profile, we use the spatially-resolved $\Upsilon_*$ measurements of \cite{vanDokkum+2017}, shown in Fig.~\ref{fig:ml_i}.  Details of the Low Resolution Imaging Spectrograph (LRIS) observations, data reduction, and modeling can be found in Sections 2 and 3 of \cite{vanDokkum+2017}.  Fitting of the extracted 1D spectra was performed with the stellar population synthesis (SPS) models of \citet{conroy+2017b} (an update to those of \citealt{conroy&vanDokkum2012b}), using the extended stellar library of \cite{villaume+2017} and the MIST stellar isochrones \citep{choi+2016}.  The logarithmic slope of the initial mass function (IMF) was allowed to vary in the ranges of $0.08 < M / M_\odot < 0.5$ and $0.5 < M / M_\odot < 1$.  For $M / M_\odot > 1$, a \cite{salpeter1955} log slope of $-2.35$ was adopted.

\begin{figure}
  \centering
  \includegraphics[width=\linewidth]{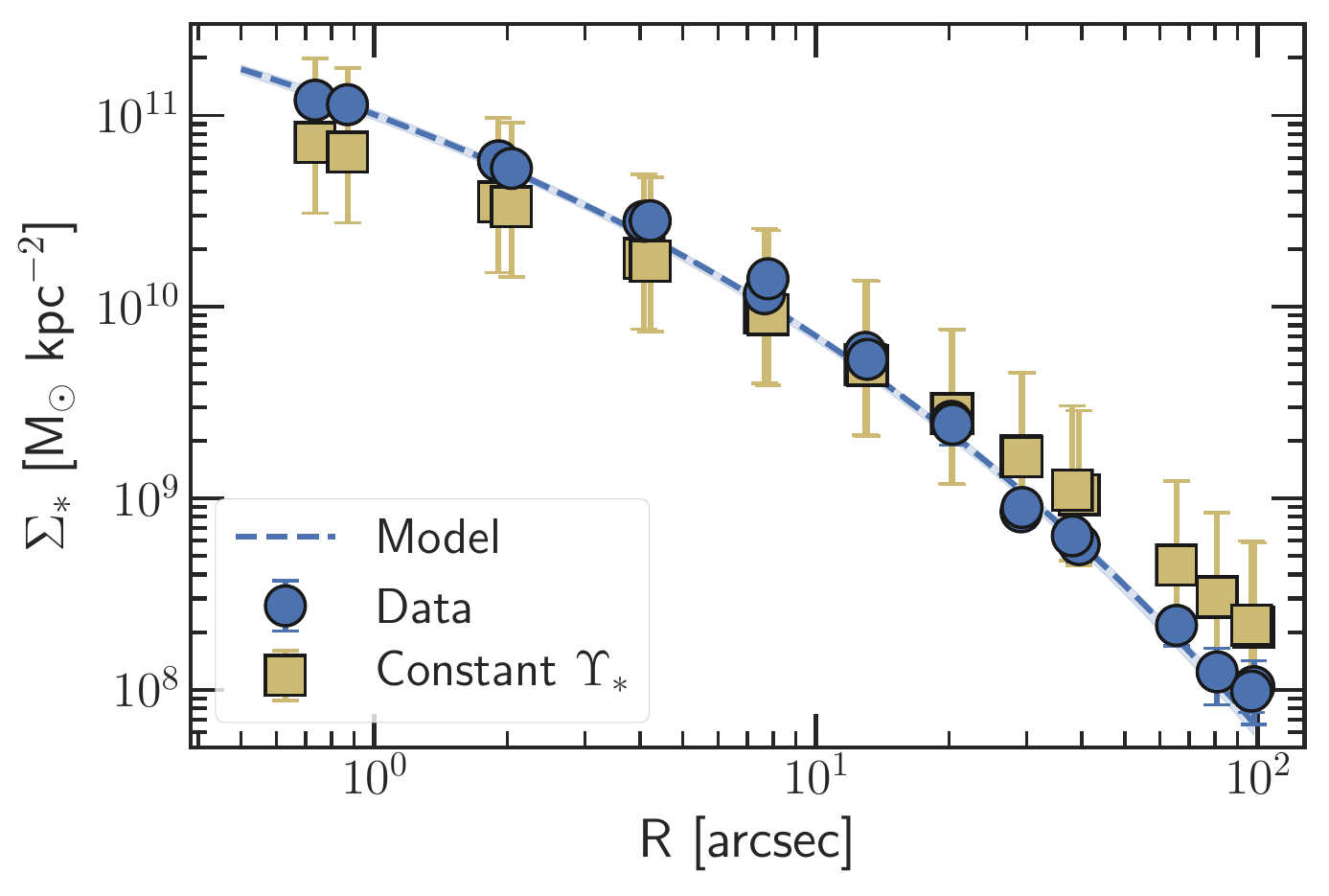}
  \caption{Stellar mass surface density.  Blue circles show the measured values using the variable $\Upsilon_*$ profile.  The blue dashed line shows the best fit model (see Sec.~\ref{sec:methods}) of the surface density, with the width of the curve showing the inner 68\% of samples.  We compare this with a profile derived from a constant $\Upsilon_*$, shown as the yellow squares.  The uncertainties on these points are taken from the typical uncertainties on $\Upsilon_*$.}
  \label{fig:stellar_mass_surface_density}
\end{figure}

Since these $\Upsilon_*$ values were computed for $I$ band, we use the $B - I$ color profile measured by \cite{spolaor+2008a} to convert to a $B$ band $\Upsilon_*$.  We then multiply these $\Upsilon_*$ measurements by the stellar surface brightness profile to obtain the mass surface density profile shown in Fig.~\ref{fig:stellar_mass_surface_density}, propagating uncertainties under the assumption that the $\Upsilon_*$ uncertainties dominate over the photometric uncertainties.  We compare this variable $\Upsilon_*$ density profile with one determined from multiplying the surface brightness profile by a constant $\Upsilon_* = 8.61$ (chosen to match the two enclosed stellar mass values at $100\arcsec$).  We see that the variable $\Upsilon_*$ profile is noticeably more compact than the constant $\Upsilon_*$ profile.  We discuss this more in Sec.~\ref{sec:discuss_ml}.

The \sersic{} fits to the stellar luminosity and mass surface density profiles are listed in Table~\ref{tab:density}.

\subsection{GC density}
\label{sec:gcdens}

Nearly all massive ETGs have been found to have GC systems with a bimodal color distribution \citep{brodie&strader2006}, and NGC~1407 is no exception \citep{forbes+2006}.  The red and blue modes are expected to trace metal-rich and metal-poor GCs respectively, with the basic galaxy formation scenario associating metal-rich GCs with in-situ star formation and metal-poor GCs with accretion \citep{brodie&strader2006, peng+2006, harris+2017}.

\begin{figure}
  \centering
  \includegraphics[width=\linewidth]{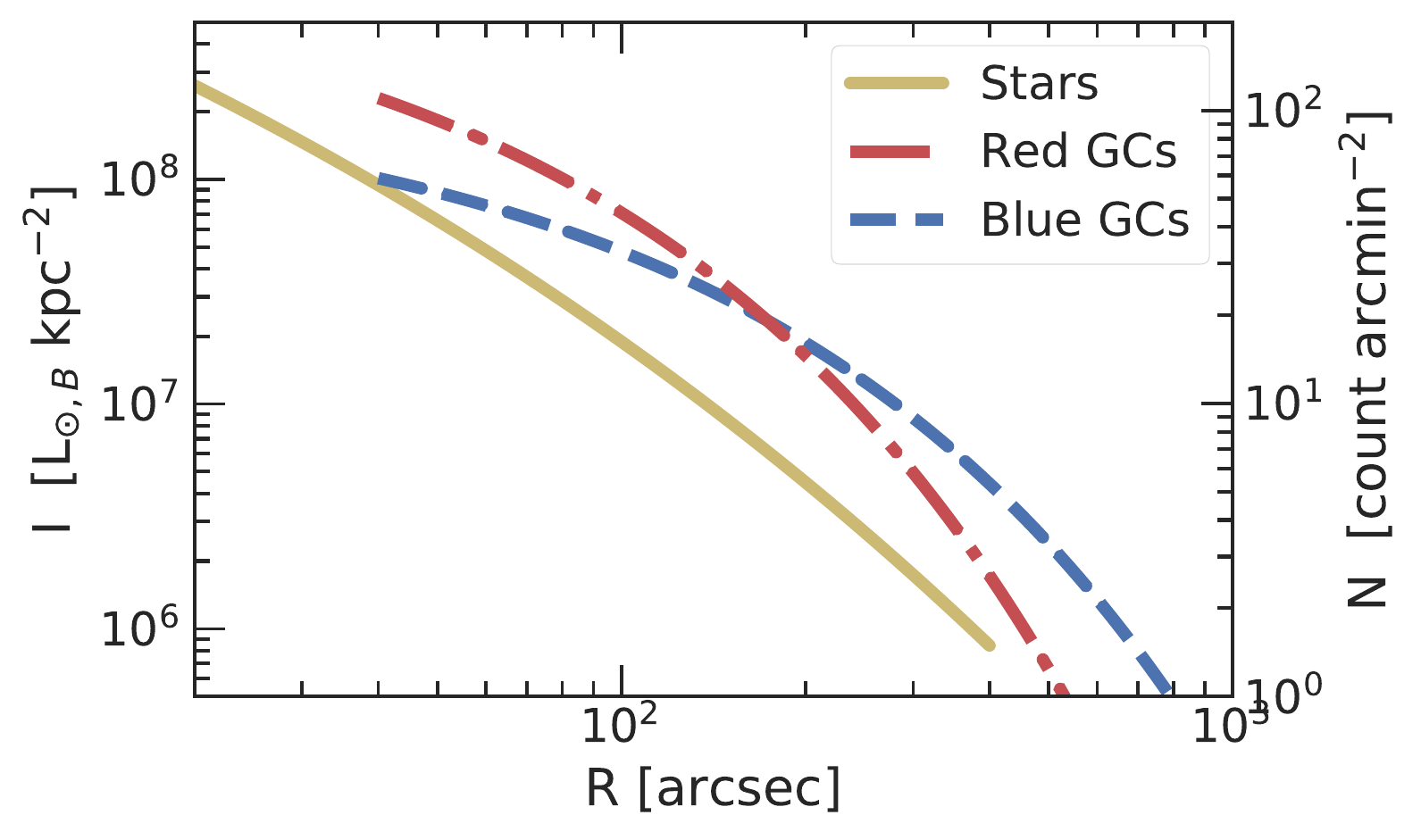}
  \caption{Surface brightness and surface number density profiles for the field stars, blue GCs, and red GCs.  The extent of the radial ranges represent where the profiles were fitted to the photometric data.}
  \label{fig:surface_density}
\end{figure}

Since we model the dynamics of the blue and red GC subpopulations simultaneously, we use separate surface number density profiles for each subpopulation, using the results from \cite{pota+2013}.  With Subaru/Suprime-Cam $g$ and $i$ band imaging, they fitted a single \sersic{} profile plus uniform background contamination model to both the red and blue subpopulations, splitting the two subpopulations at a color of $g - i = 0.98$ mag.  Their resulting \sersic{} parameters are listed in Table~\ref{tab:density}, and the profiles are shown in Fig.~\ref{fig:surface_density}.

Relative to the field star density distribution, the GC profiles show flatter inner cores, possibly due to tidal destruction of GCs at small galactocentric radii.  The red GC subpopulation is more compact than the blue GCs, though both are far more spatially extended than the field stars.  

\begin{figure}
  \centering
  \includegraphics[width=\linewidth]{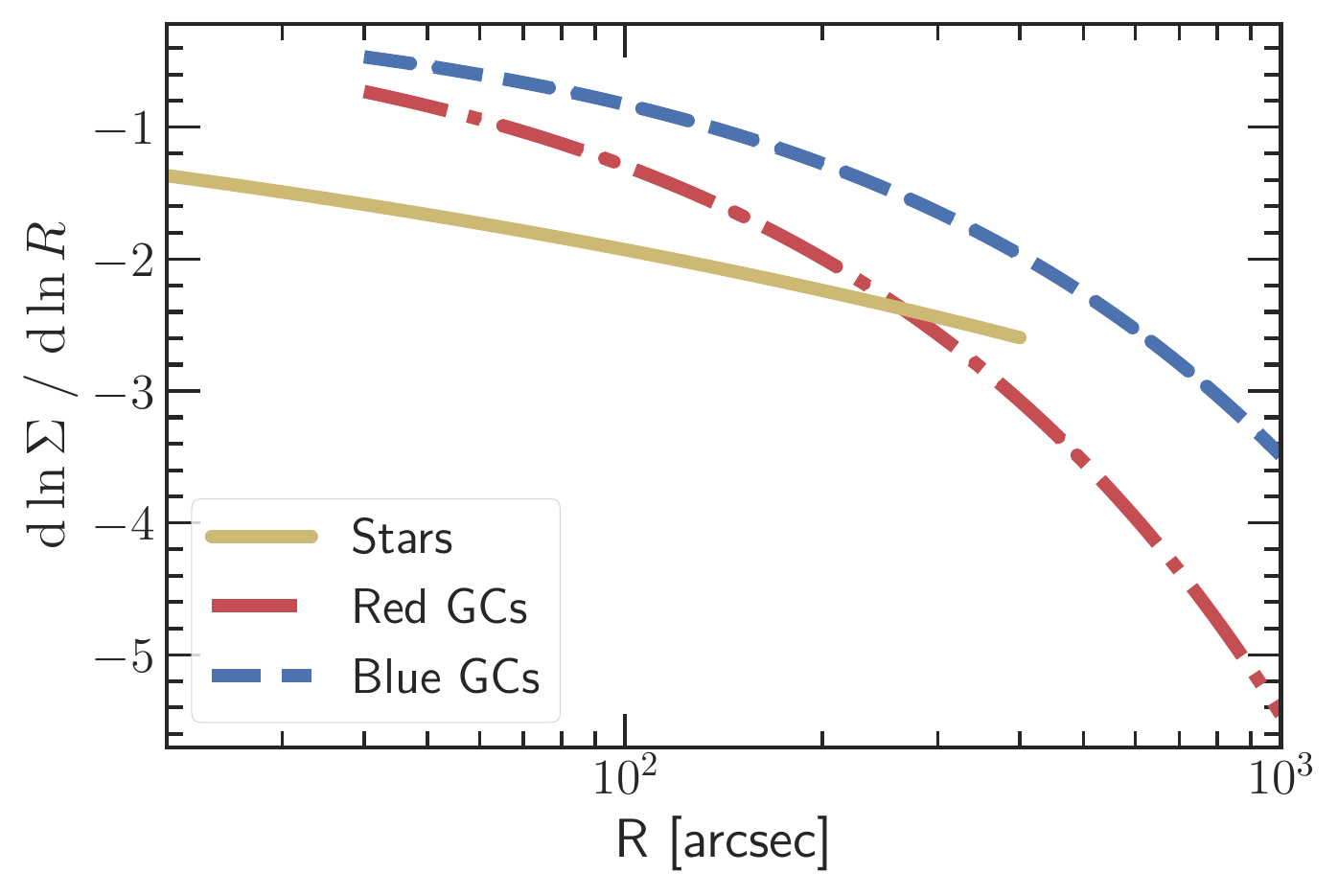}
  \caption{Log-slopes of the surface brightness and surface number density profiles for the field stars, blue GCs, and red GCs.  The extent of the radial ranges represent where the profiles were fitted to the photometric data.}
  \label{fig:density_slopes}
\end{figure}

In Fig.~\ref{fig:density_slopes} we show the log-slopes of the tracer surface density profiles as a function of radius.  The density slope of the red GC subpopulation qualitatively matches that of the field stars in the outer halo, matching expectations that the metal-rich GCs are associated with the field star population \citep{forbes+2012}.

\begin{deluxetable}{llll}
  \tablecaption{\sersic{} profile parameters\label{tab:density}}
  \tablehead{& \colhead{$I_0$} & \colhead{$\Reff$} & \colhead{$n$}}
  \startdata
  Stellar luminosity & 1.55\E{11} & $100\arcsec \pm 3\arcsec$ & $4.67 \pm 0.15$ \\
  Stellar mass & 3.25\E{12} & $23\arcsec \pm 2\arcsec$ & $3.93 \pm 0.05$ \\
  Red GCs & 354 & $169\arcsec \pm 7\arcsec$ & $1.6 \pm 0.2$ \\
  Blue GCs & 124  & $346\arcsec \pm 30\arcsec$ & $1.6 \pm 0.2$ \\
  \enddata
  \tablecomments{Left to right: central surface density, effective radius, and \sersic{} index. The central surface density has units of L$_{\odot, B}$ kpc$^{-2}$ for the stellar luminosity, M$_\odot$ kpc$^{-2}$ for the stellar mass, and count arcmin$^{-2}$ for the GCs. \todo{check stellar mass $I_0$}}
\end{deluxetable}

\subsection{Stellar kinematics}
\label{sec:stkin}
In the inner $\sim 40\arcsec$ (0.4 \Reff{}) of the galaxy, we use longslit spectroscopy along the major axis from the ESO Faint Object Spectrograph and Camera (EFOSC2), originally analyzed by \cite{spolaor+2008a}.  These data were re-analyzed by \cite{proctor+2009}, who used penalized pixel fitting \citep{cappellari&emsellem2004} to calculate a velocity dispersion profile for the galaxy.  

Here we define the velocity dispersion as the root mean square (RMS) velocity,   $v_\mathrm{rms} = \sqrt{\langle v^2 \rangle}$.  For the longslit data along the major axis, we account for the slight rotational motion by calculating $v_\mathrm{rms}$ as
\begin{equation}
  v_\mathrm{rms} = \sqrt{\frac{v_\mathrm{rot}^2}{2} + \sigma^2}  
\end{equation}
where $\sigma$ is the standard deviation of the line-of-sight velocity distribution (LOSVD).

\begin{figure}
  \centering
  \includegraphics[width=\linewidth]{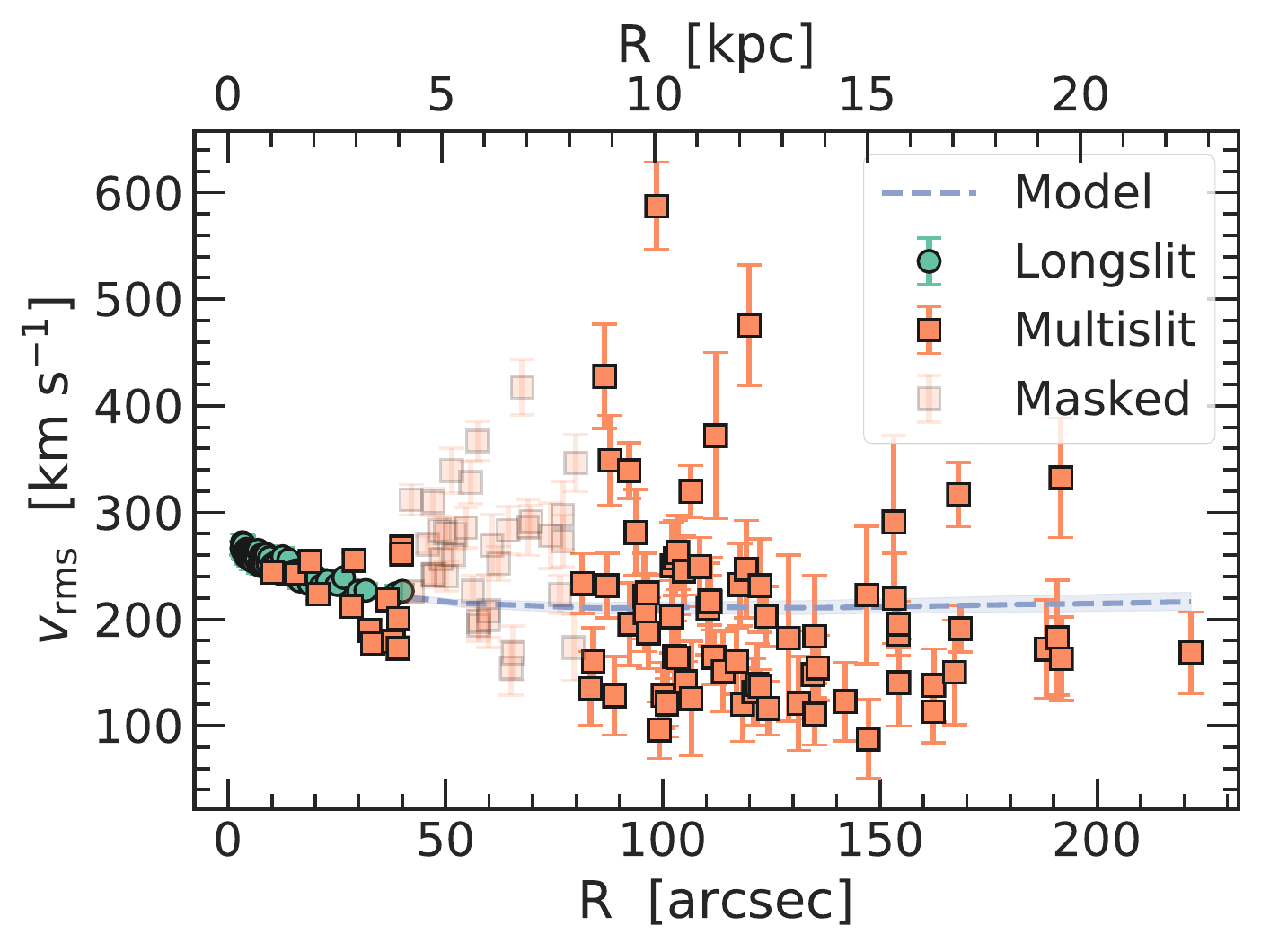}
  \caption{Stellar velocity dispersion data out to 2 \Reff{} ($\sim 20$ kpc).  The lighter points between $40\arcsec$ and $80\arcsec$ show where we mask data due to substructure.  The green circles show the longslit data, the orange squares show the multislit data, and the blue line shows the best fit stellar velocity dispersion model described in Sec.~\ref{sec:methods}, with the width of the curve showing the inner 68\% of samples.  We note that the error bars have {\it not} been visually scaled following the best-fit weighting parameter (Sec.~\ref{sec:measurement}).}
  \label{fig:stkin}
\end{figure}

To reach out to much farther galactocentric radii, we use the Keck/DEIMOS multislit observations presented by \cite{arnold+2014} and \cite{foster+2016}, which sample the stellar light in 2D.  Using only spectra visually classified as ``good'' by \cite{foster+2016}, these stellar velocity dispersion measurements reach out to $\sim 200\arcsec$ ($2 \Reff$), though of course with sparser spatial sampling than the longslit kinematic data.  We calculate the velocity dispersion for these 2D measurements as
\begin{equation}
  v_\mathrm{rms} = \sqrt{v_\mathrm{rot}^2 + \sigma^2} \ .
\end{equation}
These stellar kinematic measurements are shown in Fig~\ref{fig:stkin}.  

There are two complications in pre-processing the stellar kinematic data.  The first is the potential presence of substructure in the kinematics in the region between $40\arcsec$ and $80\arcsec$.  This deviation from a monotonically decreasing velocity dispersion profile was also seen in the velocity dispersion profile measured by \cite{vanDokkum+2017}, and it is further mirrored in the metallicity bump seen by \cite{pastorello+2014}.  Following \cite{pota+2015b}, we mask out this region for our analysis (the lighter points in Fig.~\ref{fig:stkin}). The second complication is the influence of the central super-massive black hole (SMBH).  \cite{rusli+2013} inferred the presence of a $\sim 4\E{9}$ M$_\odot$ SMBH in NGC~1407 with a corresponding sphere of influence with radius $\sim 2\arcsec$.  To avoid having to model the dynamical effects of the SMBH, we restrict our analysis to radii outside of $3\arcsec$.

\todo{compare with Veale+2017}

\subsection{GC kinematics and colors}
\label{sec:gckin}
\begin{figure}
  \centering
  \includegraphics[width=\linewidth]{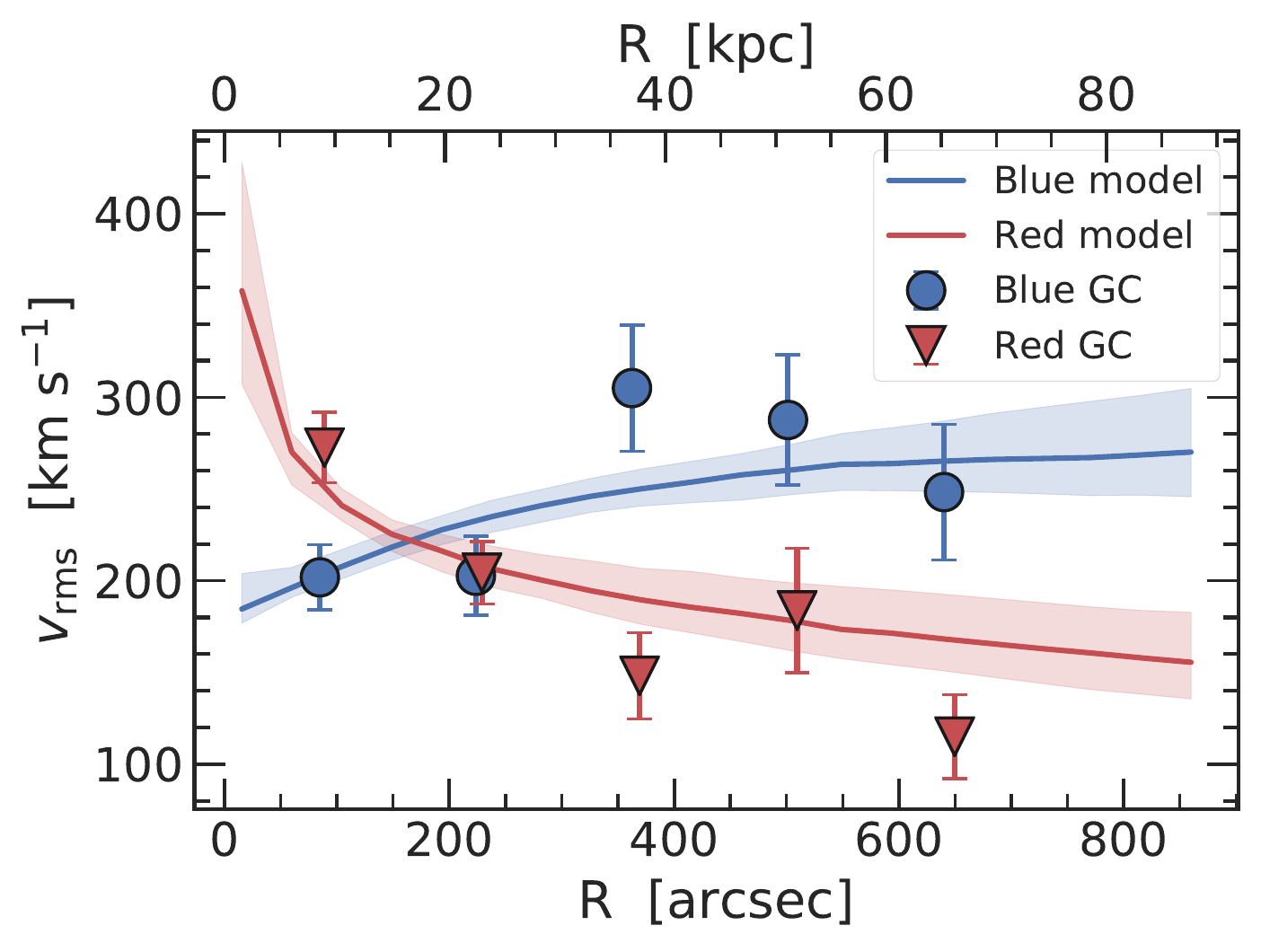}
  \caption{GC velocity dispersion vs.\ galactocentric radius, with blue GCs and red GCs showing systematically different trends in their scatter at large galactocentric radii.  The blue GC subpopulation has a rising velocity dispersion profile, while the red GC subpopulation has a falling velocity dispersion profile.  The associated best fit models of the GC velocity dispersion profiles are described in Sec.~\ref{sec:methods}, with the width of the curves showing the inner 68\% of samples.}
  \label{fig:gckin}
\end{figure}

We use the GC kinematics presented in \cite{pota+2015b}. The spectra for these measurements were obtained from ten Keck/DEIMOS slitmasks.  \todo{discuss RV outlier rejection} The red and blue GC radial velocities (RVs) in Fig.~\ref{fig:gckin} reveal that the two subpopulations have systematically different velocity dispersions in the outer regions. The GC radial velocity measurements for NGC~1407, as well as for the entire SLUGGS sample, can be found in \cite{forbes+2017c}.

\begin{figure}
  \centering
  \includegraphics[width=\linewidth]{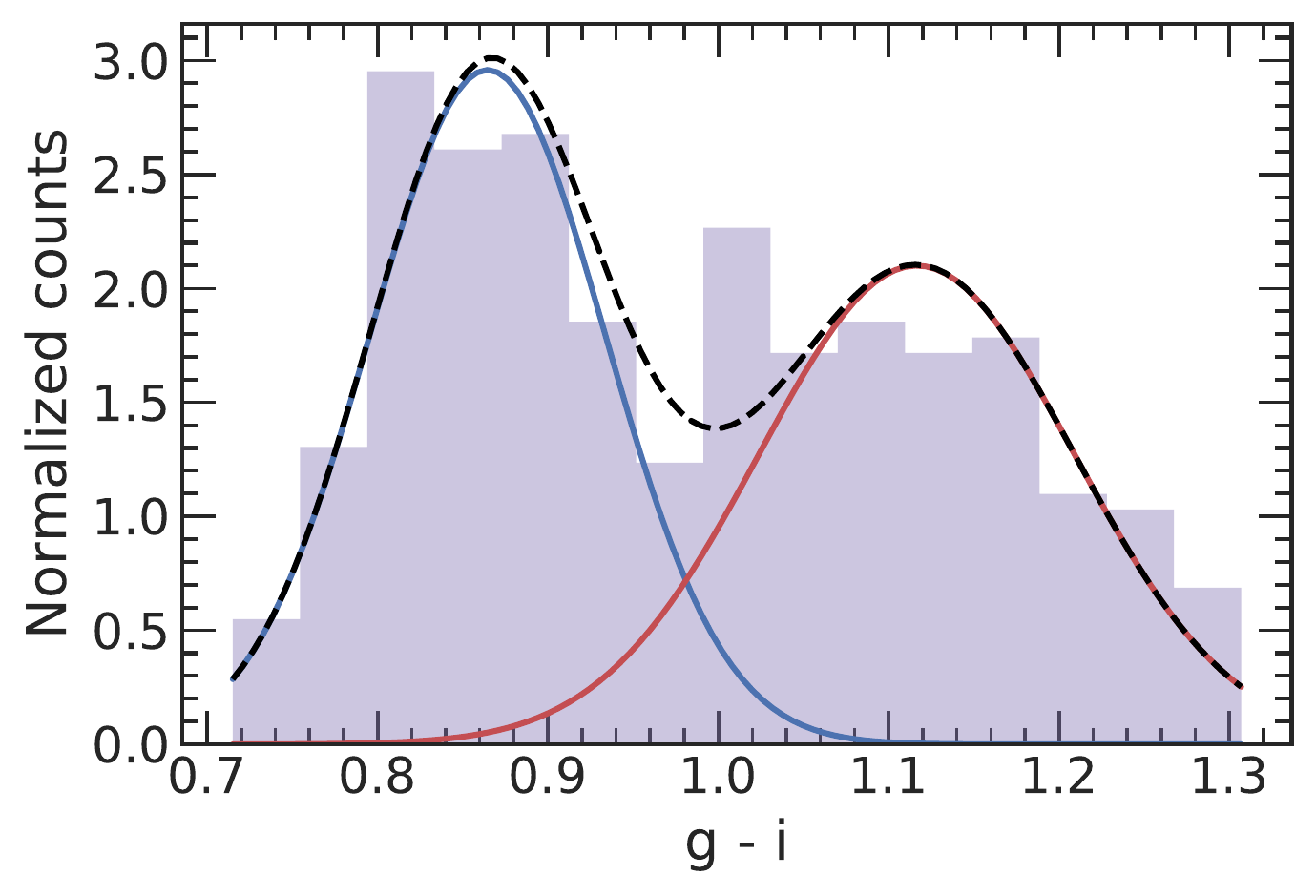}
  \caption{Gaussian mixture model of GC colors from our RV GC dataset.  The blue and red curves show the Gaussian color distribution for the blue GC and red GC subpopulations respectively, while the dashed black curve shows that the sum of these distributions accurately captures the observed RV GC color distribution (in the violet histogram).}
  \label{fig:gcgmm}
\end{figure}

The $g - i$ color distribution of our spectroscopic GC dataset, shown in Fig.~\ref{fig:gcgmm}, is well-matched to the photometric GC catalog presented in \cite{pota+2013}.  We fit a Gaussian Mixture Model to the spectroscopic sample color distribution and compare the result with the distributions found for the photometric sample of \cite{pota+2013}\footnote{We note that the mean color of the red GCs was swapped with that of NGC~2768 in the presentation of their Table 3.}, listed in Table~\ref{tab:gmm}.  We find that the color Gaussians of the RV GC sample have nearly identical means to those of the photometric sample, though the blue Gaussian of the RV sample has a slightly larger standard deviation than that of the photometric sample.

We emphasize that we do {\it not} split the GCs into red and blue subpopulations based on color for the dynamical analysis, but rather use this information to assign a probability of being in either subpopulation for each GC (Sec.~\ref{sec:measurement}).

\begin{deluxetable}{llll}
  \tablecaption{GC color Gaussian parameters \label{tab:gmm}}
  \tablehead{& \colhead{$\mu_c$} & \colhead{$\sigma_c$} & \colhead{$\phi$}}
  \startdata
  Blue GCs (photometric sample) & 0.85 & 0.05 & ---\\
  Red GCs (photometric sample) & 1.10 & 0.1 & --- \\ 
  Blue GCs (RV sample) & 0.87 & 0.067 & 0.52 \\
  Red GCs (RV sample) & 1.12 & 0.094 & 0.48 \\
  \enddata
  \tablecomments{Comparison of the color distribution of our spectroscopic GC sample with that of the GC system overall. The weights, $\phi$, indicate the fraction of GCs which come from the specified subpopulation.}
\end{deluxetable}

\section{Methods}
\label{sec:methods}

Here we describe the dynamical (Sec.~\ref{sec:dynamics}) and statistical (Sec.~\ref{sec:measurement}) methods that we use to model our data.

\subsection{Dynamical model}
\label{sec:dynamics}
Given the low $v / \sigma$ and near-circular isophotes of the galaxy, we assume spherical symmetry for our model.  Further assuming that we have a perfectly collisionless tracer population in steady-state, we can write the spherically symmetric Jeans equation as
\begin{equation}
  \label{eqn:jeans}
  \frac{\de(\nu \bar{v_r^2})}{\de r} + 2 \frac{\beta}{r}\nu \bar{v_r^2} = -\nu \frac{\de\Phi}{\de r}
\end{equation}
where $\nu$ is the volume density of the tracer, and 
\begin{equation}
  \label{eqn:beta}
  \beta \equiv 1 - \frac{\sigma_\theta^2 + \sigma_\phi^2}{2\sigma_r^2}
\end{equation}
is the standard orbital anisotropy parameter \citep{binney&tremaine2008}.

We can integrate once to obtain the mean square of the radial component of the velocity, and again to obtain the projected line-of-sight (LOS) RMS velocity.  Following \cite{mamon&lokas2005b} the latter is,
\begin{equation}
  \sigma_\mathrm{los}(R) = \frac{2G}{I(R)} \int_R^\infty K\left(\frac{r}{R}, \ \beta\right) \nu(r) M(<r) \frac{dr}{r}
\end{equation}
where $I(R)$ is the surface density profile, $\nu(r)$ is the volume density profile, $M(<r)$ is the enclosed mass profile, and $K(u, \beta)$ is the appropriate Jeans kernel.  The Jeans kernel weighs the impact of the orbital anisotropy across the various deprojected radii, $r$, associated with the projected radius, $R$.

We note that we only model the LOS $v_\mathrm{rms}$, and not any higher-order moments of the LOS velocity distribution.  If we assume that the anisotropy parameter of a tracer is constant at all radii, then we have
\begin{equation}
  \begin{split}
    K(u, \beta) = \frac{1}{2} u^{2\beta - 1} & \left[\left(\frac{3}{2} - \beta \right) \sqrt{\pi} \frac{\Gamma(\beta - 1/2)}{\Gamma(\beta)} \right. \\
    & \left. + \beta B \left(\beta + \frac{1}{2}, \frac{1}{2}; \frac{1}{u^{2}}\right) \right. \\
    & \left. - B \left(\beta - \frac{1}{2}, \frac{1}{2}; \frac{1}{u^{2}}\right) \vphantom{\frac{1}{1}} \right] 
  \end{split}
\end{equation}
where $B(a, b; z)$ is the incomplete Beta function \cite[Appendix A]{mamon&lokas2005b}.  By writing the incomplete Beta function in terms of the hypergeometric function (\citeauthor{weisstein}),
\begin{equation}
  B(a, b; z) = a^{-1} z^a \ _2F_1[a, 1 - b, a + 1; z]
\end{equation}
we can extend this formula to values of $\beta \leq 1/2$ that would otherwise make this expression undefined.

We model our tracer density as a \sersic{} profile (Eqn.~\ref{eq:sersic}).  For the blue and red GCs, the parameters of these \sersic{} profiles are fixed to the values described in Sec.~\ref{sec:gcdens}.  For the stellar density profile, we freely vary the \sersic{} parameters to jointly constrain the stellar surface density profile shown in Sec.~\ref{sec:stdens} and the impact of the stellar mass on the kinematics.

The deprojected volume density profile in the \sersic{} model is approximated as
\begin{equation}
  \begin{split}
  \nu(r) =& \ I_0 \frac{b_n^{n(1 - p_n)}}{2R_e} \frac{\Gamma(2n)}{\Gamma((3 - p_n) n)} \\ 
          &\times \left( \frac{r}{R_e} \right)^{-p_n} \exp\left( -b_n \left(\frac{r}{R_e}\right)^{1 / n}\right)
  \end{split}
\end{equation}
where the reciprocal polynomial $p_n$ can be found by minimizing the difference with this equation and the density as computed from an inverse Abel transform of the projected surface density (see Eqn.~19 in \citealt{limaneto+1999} for an appropriate series approximation to $p_n$).

We use a generalized Navarro-Frenk-White (gNFW) dark matter density profile of the form given by Eqn.~\ref{eqn:nfw}. The enclosed mass profile of this model is found by integrating the spherically-symmetric density profile,
\begin{equation}
  \begin{split}
  M_\mathrm{DM}(<r) &= \int_0^r 4\pi r'^2 \rho(r') \dee r' \\
       &= 4\pi \rho_s \int_0^r r'^2 \left(\frac{r'}{r_s}\right)^{-\gamma} \left(1 + \frac{r'}{r_s}\right)^{\gamma - 3} \dee r'
  \end{split}
\end{equation}
Comparing this with the integral form of the hypergeometric function
\begin{equation}
  _2F_1[a, b, c; z] = \frac{1}{B(b, c-b)} \int_0^1 x^{b - 1} (1 - x)^{c - b - 1} (1 - zx)^{-a} \dee x
\end{equation}
where $B(x, y)$ is the complete Beta function, we obtain
\begin{equation}
  M_\mathrm{DM}(<r) = \frac{4\pi \rho_s r_s^3}{\omega} \left(\frac{r}{r_s}\right)^{\omega} \ _2F_1\left[\omega, \omega, \omega + 1; -\frac{r}{r_s}\right]
\end{equation}
where $\omega \equiv 3 - \gamma$.

The stellar mass is the deprojected enclosed luminosity of the \sersic{} density profile, given by
\begin{equation}
  \begin{split}
    M_*(<r) = 2\pi n \Sigma_0 \left(\frac{\Reff}{b_n^n}\right)^2 \frac{\Gamma(2n)}{\Gamma((3 - p_n)n)} \\
    \times \ \gamma\left[(3 - p_n)n, b_n \left(\frac{r}{\Reff}\right)^{1/n}\right]
  \end{split}
\end{equation}
where $\Gamma(z)$ is the complete Gamma function and $\gamma(z, x)$ is the lower incomplete Gamma function \citep{limaneto+1999}.  We note that here $\Sigma_0$ refers to the central surface mass density, not the surface brightness.

The total enclosed mass is thus given by
\begin{equation}
  M(<r) = M_\mathrm{DM}(<r) + M_*(<r) + M_\mathrm{BH}
\end{equation}
where we have included the central supermassive black hole ($M_\mathrm{BH}$) as a single point mass at $r = 0$.

\subsection{Measurement model}
\label{sec:measurement}

We construct a Bayesian hierarchical model from the previously described dynamical model that is simultaneously constrained by the longslit stellar kinematics, the multislit stellar kinematics, the GC kinematics and colors, and the stellar mass surface density measurements.  Since these data cover a range of different observations and modeling assumptions, we use the hyperparameter method of \cite{hobson+2002} to allow the properties of each dataset to determine their own relative weights.

For each dataset, we assign a parameter, $\alpha$, that scales the uncertainties on the dataset as $\delta x \rightarrow \delta x / \sqrt{\alpha}$.  This $\alpha$ parameter can be interpreted as the ``trust'' in the dataset, given the other data available and the model context in which the data are being evaluated.  This is similar to the approach adopted by \cite{oldham+2016} for balancing the contribution of stellar kinematics and GC kinematics to the overall likelihood, though we assign a weight parameter to each dataset under consideration.  We do not model the co-variance between uncertainties in the datasets (see \citealt{ma&berndsen2014} for an extension of this method to covariant uncertainties).

We note that the introduction of these weight parameters means that we need to take care in specifying the likelihood.  For a typical Gaussian likelihood, we can drop the constant $1/\delta x$ uncertainty factor (where $\delta x$ refers to the measurement uncertainty), as it does not influence our sampling of the posterior distribution.  However, the log likelihood for data, $x$, drawn from a Gaussian of mean, $\mu$, and standard deviation, $\delta x / \sqrt{\alpha}$, would now be
\begin{equation}
  \ln\mathcal{L}(x, \delta x) = -\frac{1}{2} \left( \ln \left(\frac{2\pi\delta x^2}{\alpha}\right) + \alpha \left( \frac{x - \mu}{\delta x} \right) ^ 2 \right)
\end{equation}
Thus there is now a free parameter in the first term which we cannot neglect.  The following description of our joint likelihood assumes that all uncertainties on the data have already been weighted as specified here.

For the sake of visual clarity, in this section we write all velocity dispersion quantities as $\sigma$, despite the measured velocity dispersions being given by the RMS velocity and not the standard deviation of the LOSVD.

We model the stellar velocity dispersion data, $\sigma_i \pm \delta \sigma_i$, as being drawn from a Gaussian distribution about the Jeans model prediction, $\sigma_{J}(R_i)$.
\begin{equation}
  \mathcal{L}_*(\sigma_{i}, \delta\sigma_i | \sigma_{J}(R_i)) = \frac{1}{\sqrt{2\pi\delta\sigma_i^2}} \exp\left(-\frac{(\sigma_{J}(R_i) - \sigma_i)^2}{\delta\sigma_i^2}\right)
\end{equation}
We treat both the longslit and the multislit data as measuring the same kinematic tracer (and hence $\sigma_J$ for both is calculated with the same density profile and anisotropy), but we use different weight parameters as discussed above.  When separated, we use $\mathcal{L}_\mathrm{ls}$ and $\mathcal{L}_\mathrm{ms}$ to refer to the longslit and multislit likelihoods respectively.

We model the stellar mass surface density data, $\Sigma_i \pm \delta \Sigma_i$, as being drawn from a Gaussian distribution about the proposed \sersic{} profile, $\Sigma_m(R_i)$.  
\begin{equation}
  \mathcal{L}_m(\Sigma_{i}, \delta \Sigma_i | \Sigma_m(R_i)) = \frac{1}{\sqrt{2\pi\delta \Sigma_i^2}} \exp\left(-\frac{(\Sigma_m(R_i) - \Sigma_i)^2}{\delta \Sigma_i^2}\right)
\end{equation}
This is the same \sersic{} profile used for the mass modeling, and so while the parameters of this model are primarily constrained by the data presented in Sec.~\ref{sec:stdens}, these \sersic{} parameters also influence the predicted kinematic data.

Our analysis of the GC kinematic data differs from that of \cite{pota+2015b} in that we do not use a strict color cut or bin GC RV measurements by radius.  Rather, we follow the approach of \cite{zhu+2016} in modeling GC RVs as a mixture of Gaussians associated with each GC subpopulation. Here the mean velocity is the systemic velocity of the galaxy and the standard deviation is the predicted $\sigma_J$ from the Jeans model associated with that subpopulation.  We model the GC colors as being drawn from the mixture of Gaussians as described in Sec.~\ref{sec:gckin}.

Thus the likelihood for a particular GC measurement (with velocity $v_i \pm \delta v_i$ and $g - i$ color $c_i \pm \delta c_i$), under the assumption that it comes from a particular subpopulation, $k$, (described by a distinct density profile, anisotropy, and color distribution) is
\begin{equation}
  \begin{split}
  & \mathcal{L}_k(v_i, \delta v_i, c_i, \delta c_i | \sigma_{J, k}(R_i)) = \\
  & \frac{1}{\sqrt{2\pi(\delta v_i^2 + \sigma_{J, k}^2(R_i))}} \exp\left(-\frac{v_i^2}{\delta v_i^2 + \sigma_{J, k}^2(R_i)}\right) \\
  & \times \frac{1}{\sqrt{2\pi(\delta c_i^2 + \sigma_{c, k}^2)}} \exp\left(-\frac{(c_i - \mu_{c, k})^2}{\delta c_i^2 + \sigma_{c, k}^2}\right)
  \end{split}
\end{equation}
where $\mu_{c, k}$ and $\sigma_{c, k}$ are the mean and standard deviation of the color Gaussian for the $k$-th subpopulation, and $\sigma_{J, k}$ is the Jeans model prediction. 

The likelihood for the GC data is therefore
\begin{equation}
  \mathcal{L}_\mathrm{gc}(v_i, \delta v_i, c_i, \delta c_i) = \sum_{k \in \{b, r\}} \phi_k \mathcal{L}_k(v_i, \delta v_i, c_i, \delta c_i)
\end{equation}
where $\phi_k$ is the mixture model weight for the $k$-th GC subpopulation, satisfying $\sum_k \phi_k = 1$.  We note that the probability that an individual GC comes from a particular subpopulation is given by
\begin{equation}
  P_k(v, \delta v, c, \delta c) = \frac{\phi_k \mathcal{L}_k(v, \delta v, c, \delta c)}{\sum_j \phi_j \mathcal{L}_j(v, \delta v, c, \delta c)} \ .
\end{equation}

Putting all of the likelihoods together, our final joint likelihood is
\begin{equation}
  \mathcal{L} = \prod_i \mathcal{L}_\mathrm{ls} \times \prod_i \mathcal{L}_\mathrm{ms} \times \prod_i \mathcal{L}_\mathrm{m} \times \prod_i \mathcal{L}_\mathrm{gc} \ .
\end{equation}
In practice, we compute the log-likelihood.
\begin{equation}
  \ln \mathcal{L} = \ln\mathcal{L}_\mathrm{ls} + \ln\mathcal{L}_\mathrm{ms} + \ln\mathcal{L}_\mathrm{m} + \ln\mathcal{L}_\mathrm{gc} 
\end{equation}

Our model has fifteen free parameters, listed in Table~\ref{tab:params}.  The parameters are as follows: the scale density of the DM halo ($\rho_s$), the scale radius of the DM halo ($r_s$), the inner DM density log-slope ($\gamma$), the SMBH mass ($M_\mathrm{bh}$), the anisotropy of the field stars ($\beta_s$), the anisotropy of the blue GCs ($\beta_b$), the anisotropy of the red GCs ($\beta_r$), the distance ($D$), the central stellar mass surface density ($\Sigma_{0,*}$), the stellar mass effective radius ($\Reff$), the stellar mass \sersic{} index ($n_*$), the weight for the longslit dataset ($\alpha_\mathrm{ls}$), the weight for the multislit dataset ($\alpha_\mathrm{ms}$), the weight for the GC dataset ($\alpha_\mathrm{gc}$), and the weight for the stellar mass surface density dataset ($\alpha_\mathrm{m}$).

The dynamical model and measurement model was constructed with {\sc slomo}\footnote{\url{https://github.com/adwasser/slomo}}, a \emph{python}-based code doing Jeans modeling of spherically symmetric systems.  To sample our posterior probability distribution, we use \emph{emcee} \citep{emcee2013}, an implementation of the affine-invariant Markov Chain Monte Carlo (MCMC) ensemble sampler described by \cite{goodman&weare2010}.  We run our sampler with 128 walkers for 6000 iterations, rejecting the first 4500 iterations where the chains have not yet fully mixed.  The traces of these walkers are shown in Appendix~\ref{sec:post} in Fig.~\ref{fig:walkers}.

\begin{deluxetable}{llll}
  \tablecaption{Model parameters \label{tab:params}}
  \tablehead{\colhead{Parameter} & \colhead{Unit} & \colhead{Prior} & \colhead{Fit value}}
  \startdata
  $\log_{10} \rho_s$ & [M$_\odot$ kpc$^{-3}$] & $\mathcal{U}$(5.0, 9.0) & $6.71^{+0.87}_{-0.97}$ \\
  $\log_{10} r_s$ & [kpc] & $\mathcal{U}$(1.0, 3.0) & $1.75^{+0.54}_{-0.41}$ \\
  $\gamma$ & [---] & $\mathcal{U}$(0.0, 2.0) & $1.03^{+0.25}_{-0.44}$ \\
  $\log_{10}$ M$_\mathrm{BH}$ & [M$_\odot$] & $\mathcal{U}$(0.0, 11.0) & $5.03^{+3.00}_{-3.38}$ \\
  $\tilde{\beta}_s$ & [---] & $\mathcal{U}$($-$1.5, 1.0) & $-0.30^{+0.09}_{-0.10}$ \\
  $\tilde{\beta}_b$ & [---] & $\mathcal{U}$($-$1.5, 1.0) & $-1.12^{+0.33}_{-0.26}$ \\
  $\tilde{\beta}_r$ & [---] & $\mathcal{U}$($-$1.5, 1.0) & $0.20^{+0.25}_{-0.26}$ \\
  $D$ & [Mpc] & $\mathcal{N}$(26.0, 2.0) & $21.03^{+1.52}_{-1.37}$ \\
  $\log_{10} \Sigma_{0, *}$ & [M$_\odot$ kpc$^{-2}$] & $\mathcal{U}$(12.0, 13.0) & $12.52^{+0.05}_{-0.06}$ \\
  $\log_{10} R_{\mathrm{e}, *}$ & [arcsec] & $\mathcal{U}$(1.0, 2.5) & $1.41^{+0.06}_{-0.05}$ \\
  $n_*$ & [---] & $\mathcal{U}$(1.0, 8.0) & $4.07^{+0.14}_{-0.13}$ \\
  $\alpha_\mathrm{ls}$ & [---] & Exp & $1.88^{+0.42}_{-0.38}$ \\
  $\alpha_\mathrm{ms}$ & [---] & Exp & $0.13^{+0.02}_{-0.02}$ \\
  $\alpha_\mathrm{gc}$ & [---] & Exp & $1.88^{+1.58}_{-0.81}$ \\
  $\alpha_\mathrm{m}$ & [---] & Exp & $0.35^{+0.16}_{-0.11}$ \\
  \enddata
  \tablecomments{List of free parameters in our model with their best fit values. The fit values show the median of the posterior, along with the 68\% credible region.}
\end{deluxetable}

\subsection{Parameterizations and Priors}

For scale parameters such as $\rho_s$ or $r_s$, we use a uniform prior over the logarithm of the parameter.  For the anisotropy parameters, we re-parameterize to $\tilde{\beta} = -\log_{10} (1 - \beta)$.  By adopting a uniform prior over this symmetrized anisotropy parameter, we treat radial and tangential anisotropy values as equally probable.  For the distance, we adopt a Gaussian prior as discussed in Sec.~\ref{sec:data}.  In practice, we truncate this distribution for negative distances.  Following \cite{hobson+2002} we adopt an exponential prior over all weight parameters.

\subsection{Caveats}
\label{sec:caveats}

Before presenting our results, we discuss a number of caveats to our work.  We leave the relaxation of these assumptions for future work.

We have explicitly assumed that NGC~1407 has a spherically symmetric halo and stellar mass distribution.  Wet major merger remnants can produce triaxial halos, with the expectation that the stars end up in an oblate spheroid with its minor axis perpendicular to the major axis of its prolate dark matter halo \citep{novak+2006}.  However, NGC~1407 likely built up its halo through many minor mergers, and if the distribution of incoming merger orbits was largely isotropic as could be expected in a group environment, the galaxy could be expected to have a more spherical halo. \todo{citation for this last point}

The Jeans equations assume that the tracers of the potential are in equilibrium.  This requirement will be violated if there are recently accreted tracers or if the relaxation time is relatively short.  For the globular clusters in the outer halo, the long crossing times (on the order of a Gyr) ensure that the relaxation time is long, but mean that any recently accreted GCs will take a long time to phase mix.  While there is not any blatantly obvious substructure in the GC kinematic data, a quantitative description of substructure in the tracer population would require a more rigorous determination of the completeness of our kinematic sample.

We assume that the LOSVD is intrinsically Gaussian.  More detailed models (e.g., \citealt{romanowsky&kochanek2001, napolitano+2014}) would be necessary to make use of higher-order moments of the LOSVD. \todo{talk to Busola about deviation from Gaussianity in GC RVs}

Another major assumption we make is that the orbital anisotropies of our tracers are constant with radius.  Generically, we would expect the anisotropy to take on different values at different distances from the center of the galaxy (e.g., \citealt{xu+2017}).  There are a multitude of ways of parameterizing this anisotropy profile, including that presented by \cite{merritt1985} and that preferred by \cite{mamon&lokas2005b}.  However, given the diversity of anisotropy profiles seen in simulated galaxies, a non-parametric Jeans method such as that used by \cite{read&steger2017} may be preferred to capture a fuller range of possible orbital distributions.  

To the extent that we expect any cores created in DM halos to have their own spatial scale independent of the scale radius of the halo, a more robust test to distinguish between a DM cusp and a core should treat these two radii separately.  For instance, one can allow for a DM core out to some $r_\mathrm{core}$, then have $\rho \propto r^{-1}$ between that core radius and the scale radius, $r_s$, then transition to having $\rho \propto r^{-3}$ as in a standard NFW halo.
\todo{talk about Einasto halo?}

\section{Results}
\label{sec:results}

\begin{figure}
  \centering
  \includegraphics[width=\linewidth]{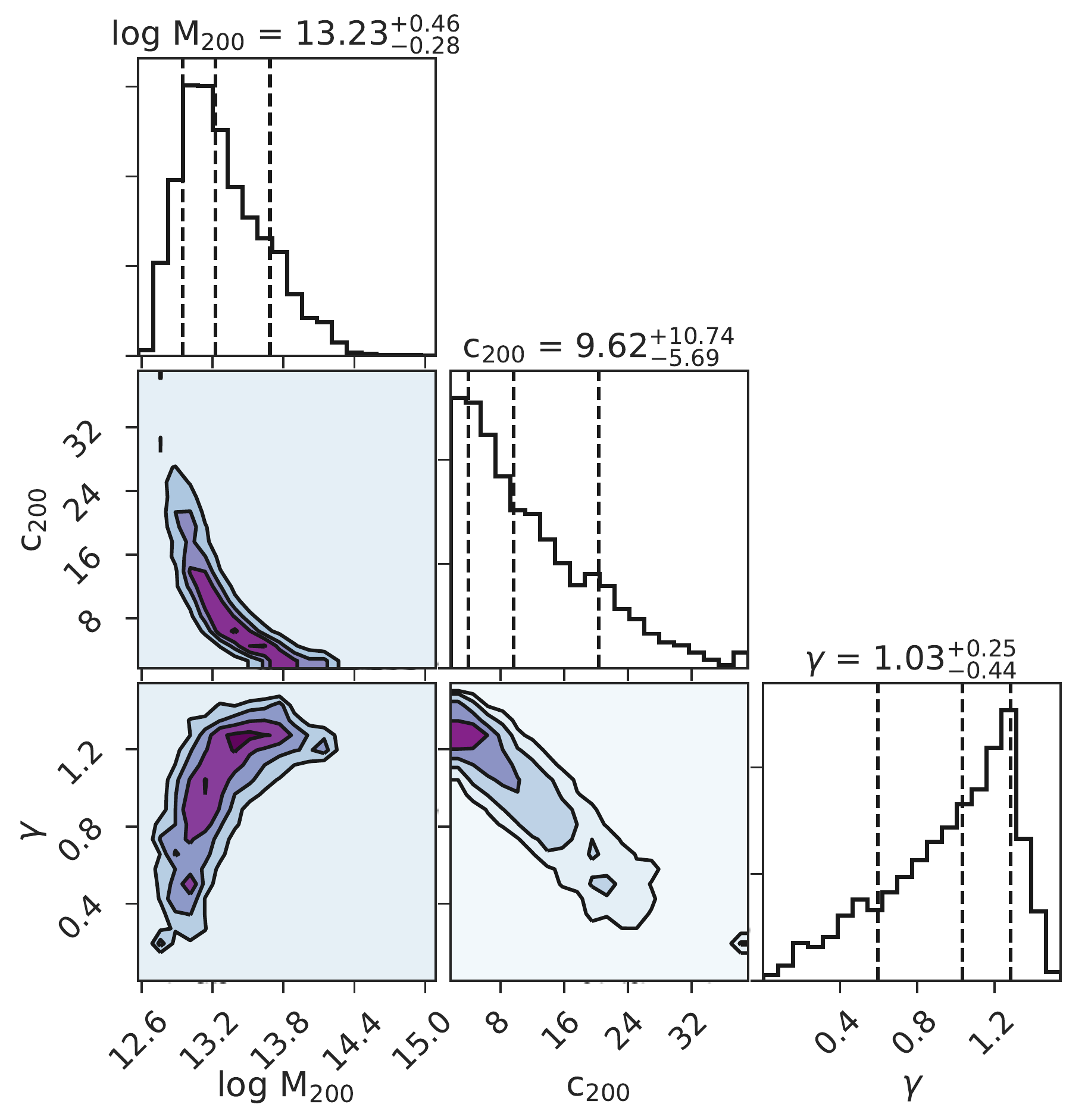}
  \caption{Posterior distribution of halo parameters.  Histograms along the diagonal show the marginalized posterior distributions of halo mass (in $M_\odot$), halo concentration, halo inner density slope.  The dashed vertical lines mark the 16th, 50th, and 84th percentiles.  The contours (at levels equivalent to 0.5, 1, 1.5 and 2 $\sigma$ for a 2D Gaussian distribution) show the covariances between these parameters.  We note the strong degeneracy among all three halo parameters.}
  \label{fig:dm_corner}
\end{figure}

We show the full posterior distribution in Fig.~\ref{fig:posterior} in Appendix~\ref{sec:post}.  We show the DM halo parameters in Fig.~\ref{fig:dm_corner}, where we have converted the halo parameters of $\rho_s$ and $r_s$ to the virial halo mass and concentration.  We use the convention of defining the virial mass as the enclosed mass with an average density $200$ times that of the critical density of the universe at $z = 0$.
\begin{equation}
  M_{200} \equiv M(<r_{200}) = \frac{4\pi}{3} r_{200}^3 \left(200\rho_\mathrm{crit}\right)
\end{equation}
The halo concentration is then defined as $c_{200} \equiv r_{200} / r_s$.  With the \cite{planck2015} cosmological parameters, $\rho_{\mathrm{crit}}(z = 0) = 127.58$ M$_\odot$ kpc$^{-3}$.  These halo parameters, along with other derived quantities, are reported in Table~\ref{tab:derived}.  We find strong evidence for a dark matter cusp in NGC~1407, with $\gamma = 1.0^{+0.3}_{-0.4}$.  The posterior distribution has 88.4\% of samples with $\gamma > 0.5$, disfavoring a cored-NFW profile.

\begin{deluxetable}{lll}
  \tablecaption{Derived parameters \label{tab:derived}}
  \tablehead{& \colhead{Unit} & \colhead{Value}}
  \startdata
  $\log_{10} M_{200}$ & [M$_\odot$] & $13.2^{+0.5}_{-0.3}$ \\
  $c_{200}$ & [---] & $10.^{+11}_{-6}$ \\
  $r_{200}$ & [kpc] & $540^{+230}_{-100}$ \\
  $\log_{10} M_*$ & [M$_\odot$] & $11.34^{+0.07}_{-0.07}$ \\
  $f_\mathrm{DM}(<5R_\mathrm{eff})$ & [---] & $0.90^{+0.01}_{-0.02}$ \\
  \enddata
  \tablecomments{List of quantities derived from free parameters. $\gamma$ is listed in Table~\ref{tab:params}.}
\end{deluxetable}

Our best fitting model predictions are shown along with the corresponding data for the stellar kinematics in Fig.~\ref{fig:stkin}, for the GC kinematics in Fig.~\ref{fig:gckin}, and for the stellar mass surface density in Fig.~\ref{fig:stellar_mass_surface_density}.

\begin{figure}
  \centering
  \includegraphics[width=\linewidth]{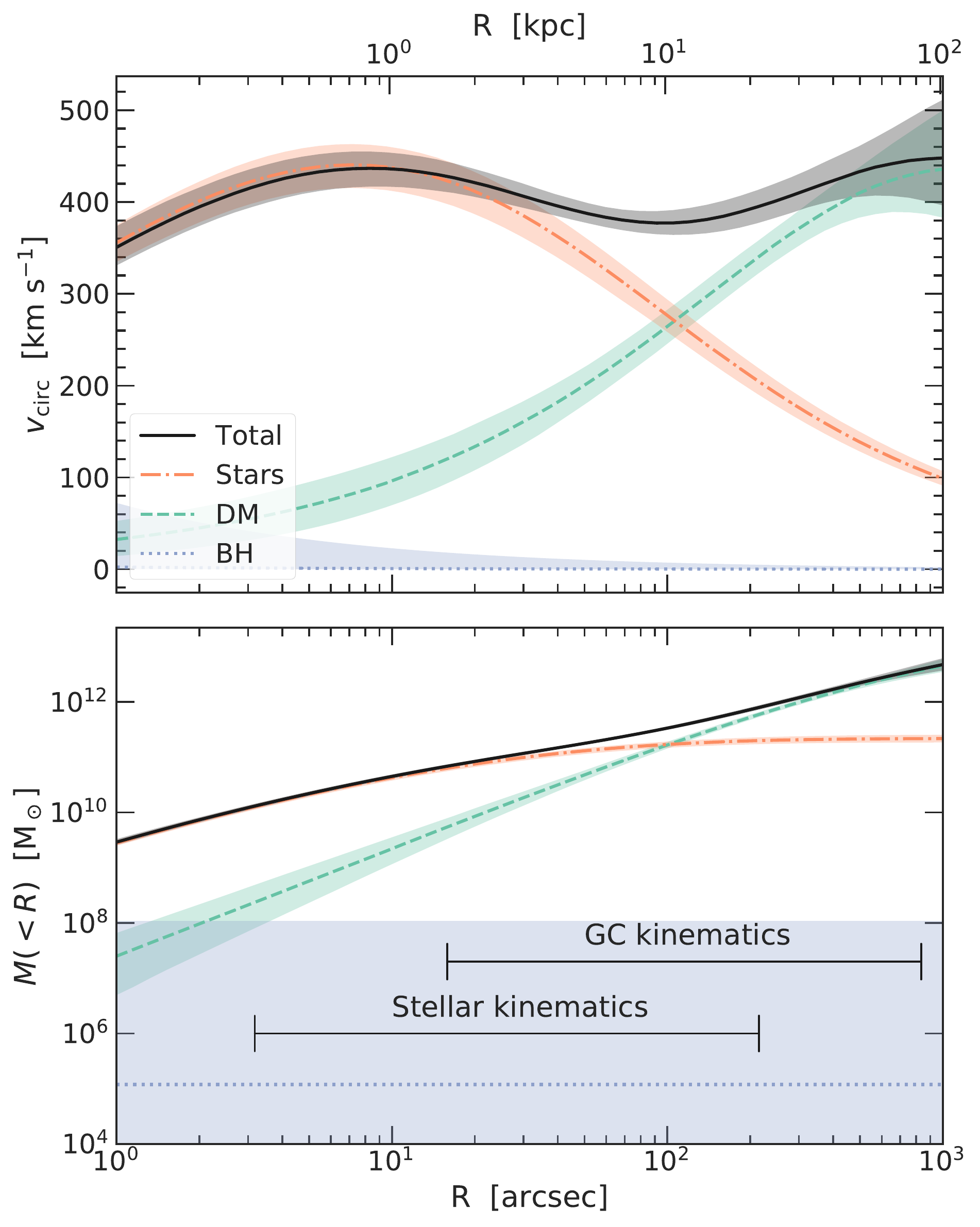}
  \caption{Top: circular velocity profiles of the mass components of the best fit model, with the width of the curve showing the central 68\% of samples.  Bottom: same, but showing the enclosed mass of each component. The horizontal bars indicate where we have constraints from stellar and GC kinematics. \todo{where would fuzzy DM bump be?}}
  \label{fig:mass_res}
\end{figure}

We show the decomposition of the enclosed mass profile into stellar, DM, and BH components in Fig.~\ref{fig:mass_res}.  Here we see that the overlap in the spatial regions probed by the GC and stellar kinematic data cover the crucial region where the DM halo becomes gravitationally dominant over the stellar mass.  As anticipated, we have weak constraints on the mass of the central SMBH, which we have treated as a nuisance parameter in the modeling.

\subsection{Mock data test}
\label{sec:mock}

Using the median model parameters shown in Table~\ref{tab:params}, we generate mock datasets to verify that our recovery of the model parameters is consistent and minimally biased.

We take the median of the posterior probability distribution as a ``true'' value.  For each stellar kinematic dataset, we use our dynamical model to generate new velocity dispersion values at each radial sample.  We sample from a Gaussian with a standard deviation taken from the associated uncertainties in original data at the respective radial points to generate the mock stellar kinematic data.  We generate mock stellar mass surface density measurements analogously.  

For the GC dataset, we create a blue GC and red GC dataset by sampling from the respective model at each radial point for which we have data.  We then assign each radial point to either be from the blue or the red subpopulation by comparing a draw from the standard uniform distribution with the $\phi_b$ value (from Table~\ref{tab:gmm}) in our model.  

When generating each dataset, we scale the standard deviation by the respective best-fit weight values (the $\alpha$ hyperparameters from Table~\ref{tab:params}).  The input uncertainties to the mock model are the same as those in the original data.

We show the recovery of our input halo model parameters in Fig.~\ref{fig:mock_dm_corner}, and we show the full parameter set in the appendix in Fig.~\ref{fig:mock}.  We find excellent recovery of the halo mass parameters.  However, our recovery of the stellar anisotropy is biased towards more tangential orbits, and our recovery of both GC subpopulations' anisotropy is biased towards more radial orbits.

\begin{figure}
  \centering
  \includegraphics[width=\linewidth]{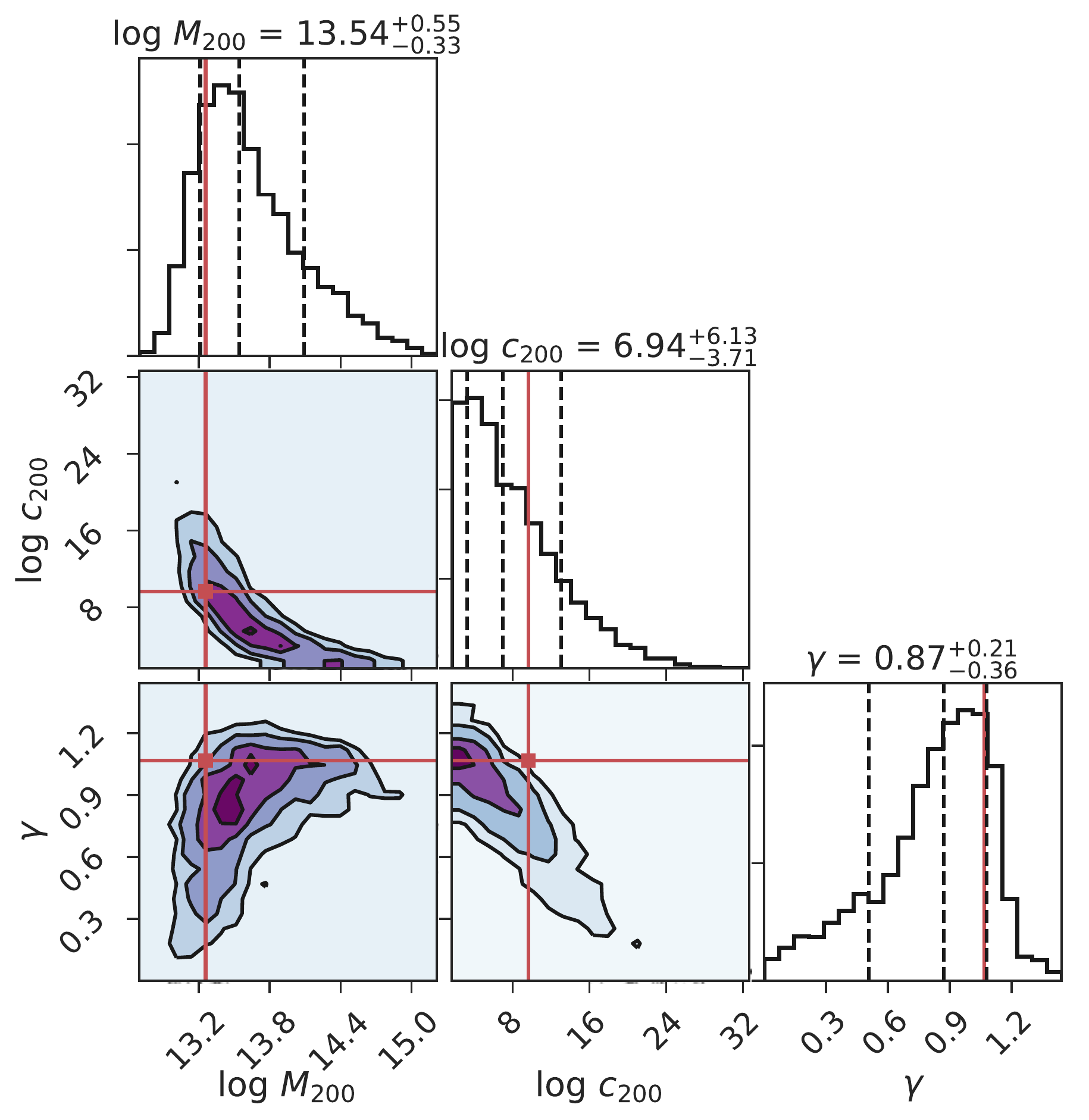}
  \caption{Posterior probability distribution from modeling our mock dataset.  Red solid lines show the model parameters used to generate the dataset.}
  \label{fig:mock_dm_corner}
\end{figure}

\subsection{Literature comparisons for NGC~1407}

\begin{figure*}
  \centering
  \subfigure{
    \includegraphics[width=0.45\linewidth]{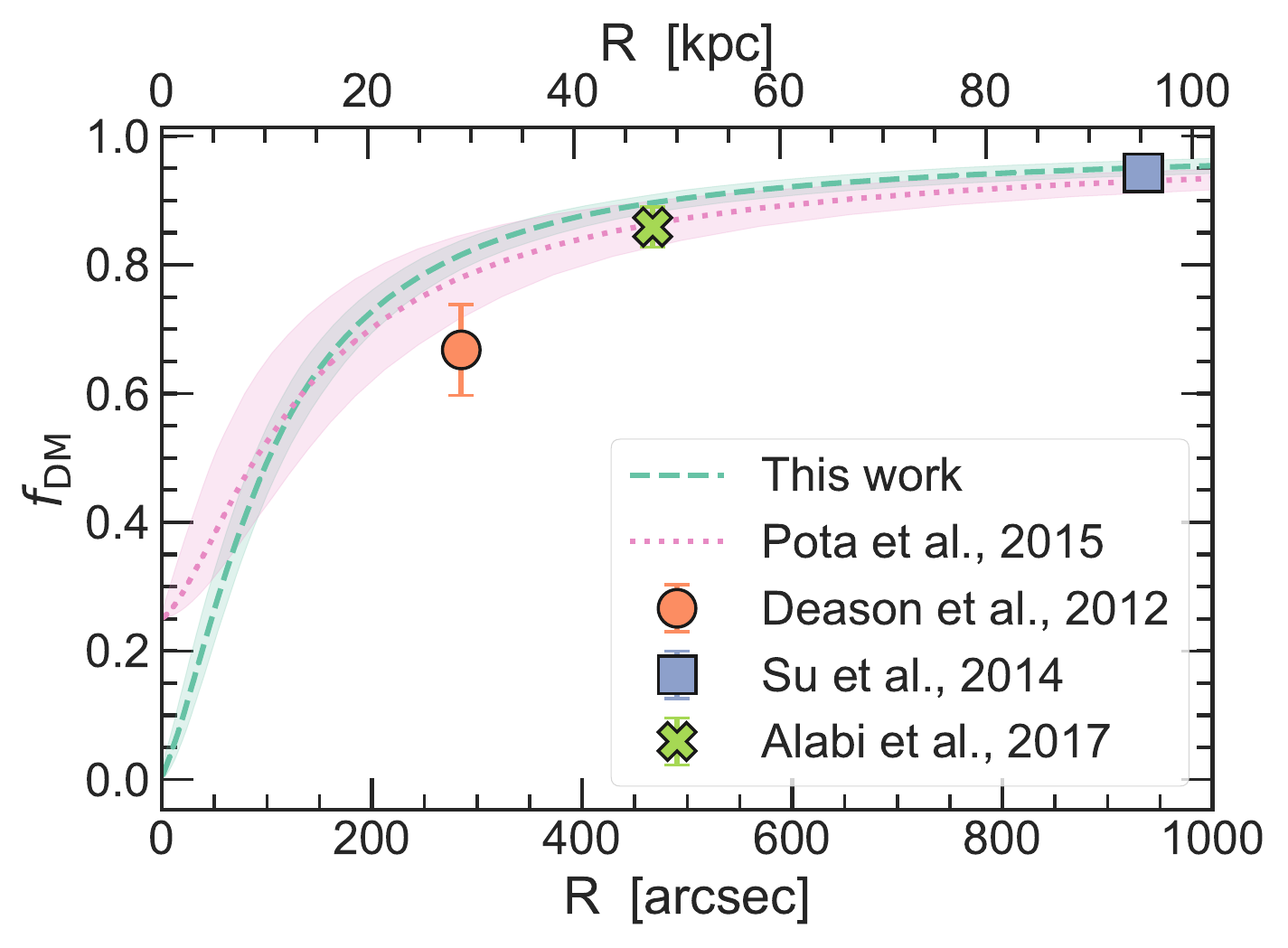}
  }
  \subfigure{
    \includegraphics[width=0.45\linewidth]{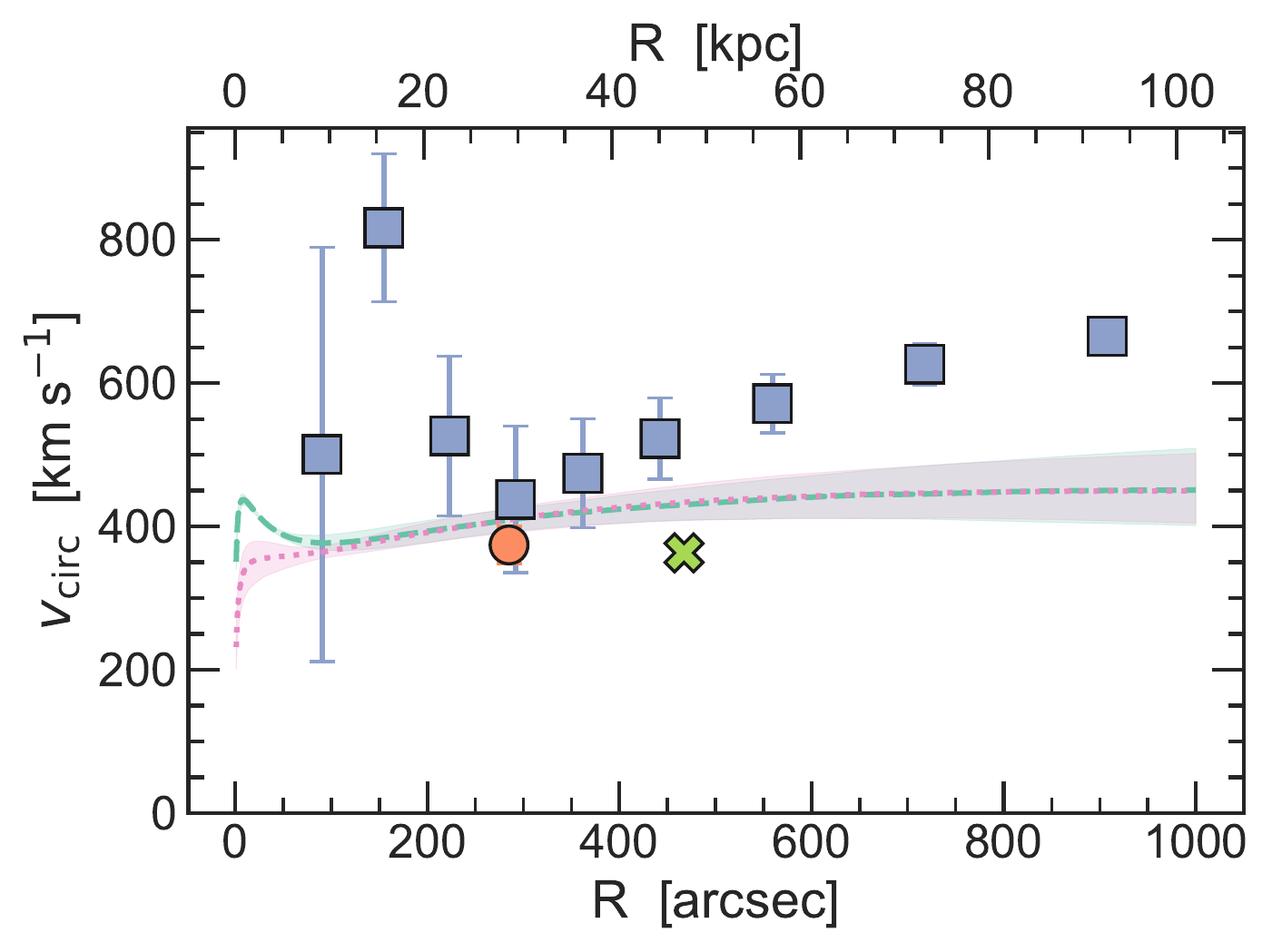}  
  }
  \caption{Left: Dark matter fraction as a function of radius, compared with measurements from the literature.  Right: Circular velocity as a function of radius, compared with measurements from the literature.  The width of the curves indicate the central 68\% of samples.}
  \label{fig:fdm_vc_lit}
\end{figure*}

In this section, we compare our mass inferences with those from some recent observational studies.  We compare both the dark matter fraction,
\begin{equation}
  f_\mathrm{DM}(<R) = 1 - M_*(<R) / M_\mathrm{tot}(<R) \ ,
\end{equation}
and the circular velocity
\begin{equation}
  v_\mathrm{circ}(R) = \sqrt{GM / R} \ .
\end{equation}
The dark matter fractions and circular velocity profiles from \cite{pota+2015b}, \cite{deason+2012}, \cite{su+2014}, and \cite{alabi+2017} are compared with our results in Fig.~\ref{fig:fdm_vc_lit}.  These are both quantities which vary with radius, so we plot values at a given angular radius on the sky to make a proper comparison.  For the dark matter fractions, we scale the reported measurements by
\begin{equation}
  1 - f_\mathrm{DM} \rightarrow \left(\frac{d_\mathrm{us}}{d_\mathrm{them}}\right) (1 - f_\mathrm{DM})
\end{equation}
to account for the differences in their adopted distances.  The stellar mass will scale with two factors of the distance for the luminosity distance dependence, and the dynamical mass will scale inversely with one factor of distance, leading to the scaling of the baryon fraction by one factor of the adopted distance.

Our total mass result is consistent with that of \cite{pota+2015b}, who adopted a distance of $28.05$ Mpc.  This is to be expected, given that we use a similar dataset and modeling technique.  They reported $f_\mathrm{DM} = 0.83^{+0.04}_{-0.04}$ at 500\arcsec, slightly below the value of $0.90_{-0.02}^{+0.01}$ that we find at the same radius. \todo{compare across profile}


\todo{compare with samurovich}

\cite{deason+2012} used a distribution function-maximum likelihood method to constrain the mass of 15 ETGs using PNe and GCs.  They assumed a distance to NGC~1407 of 20.9 Mpc, and they modeled the total mass as a power law.  For an assumed Salpeter IMF ($6 < \Upsilon_{*, B} < 10$), they found $f_\mathrm{DM} = 0.67 \pm 0.05$ within 285\arcsec, whereas we find $f_\mathrm{DM} = 0.82_{-0.03}^{+0.02}$.

\cite{su+2014} modeled the X-ray emission of hot gas surrounding NGC~1407.  Under the assumption that the gas is in hydrostatic equilibrium, they constrained the total mass profile of the galaxy and decomposed this into stellar, gas, and DM components.  They modeled the DM halo using an NFW profile, and assumed a mass-to-light ratio of $\Upsilon_\mathrm{K} = 1.17 $ M$_\odot$ / L$_{\odot, K}$ and a distance of 22.08 Mpc.  Within the inner 934\arcsec (100 kpc at their adopted distance), they found $f_\mathrm{DM} = 0.94$.  We find $f_\mathrm{DM} = 0.95_{-0.01}^{+0.01}$ within the same enclosed area.

\cite{alabi+2017} also used GCs as tracers, but applied the tracer mass estimator technique of \cite{watkins+2010} to 32 ETGs, including NGC~1407.  They assumed a distance of 26.8 Mpc.  They reported results for multiple assumptions for $\beta$, and we compare with their result ($f_\mathrm{DM} = 0.82 \pm 0.04$ at 60.7 kpc) that assumes an anisotropy of $\beta = 0$ for all the GCs (though we note that their value of $f_\mathrm{DM}$ only varies by $0.04$ between the $\beta = -0.5$ and the $\beta = 0.5$ cases). 

We find good agreement in the measured DM fractions shown in Fig.~\ref{fig:fdm_vc_lit}, though there is a slight offset between our value and that of \cite{deason+2012}.  Our total mass estimate is largely in agreement with those of other dynamical studies of NGC~1407, though we find that the X-ray mass measurements of \cite{su+2014} are noticeably larger at $R \sim 20$ kpc and also at $R \gtrsim 50$ kpc.  This is consistent with other X-ray studies of NGC 1407 (e.g., \citealt{zhang+2007, romanowsky+2009, das+2010}, though see \citealt{humphrey+2006}), and it suggests systematic differences in the X-ray and the dynamical modeling.


\subsection{Halo mass--concentration and \\ stellar mass--halo mass relations}

\begin{figure}
  \centering
  \includegraphics[width=\linewidth]{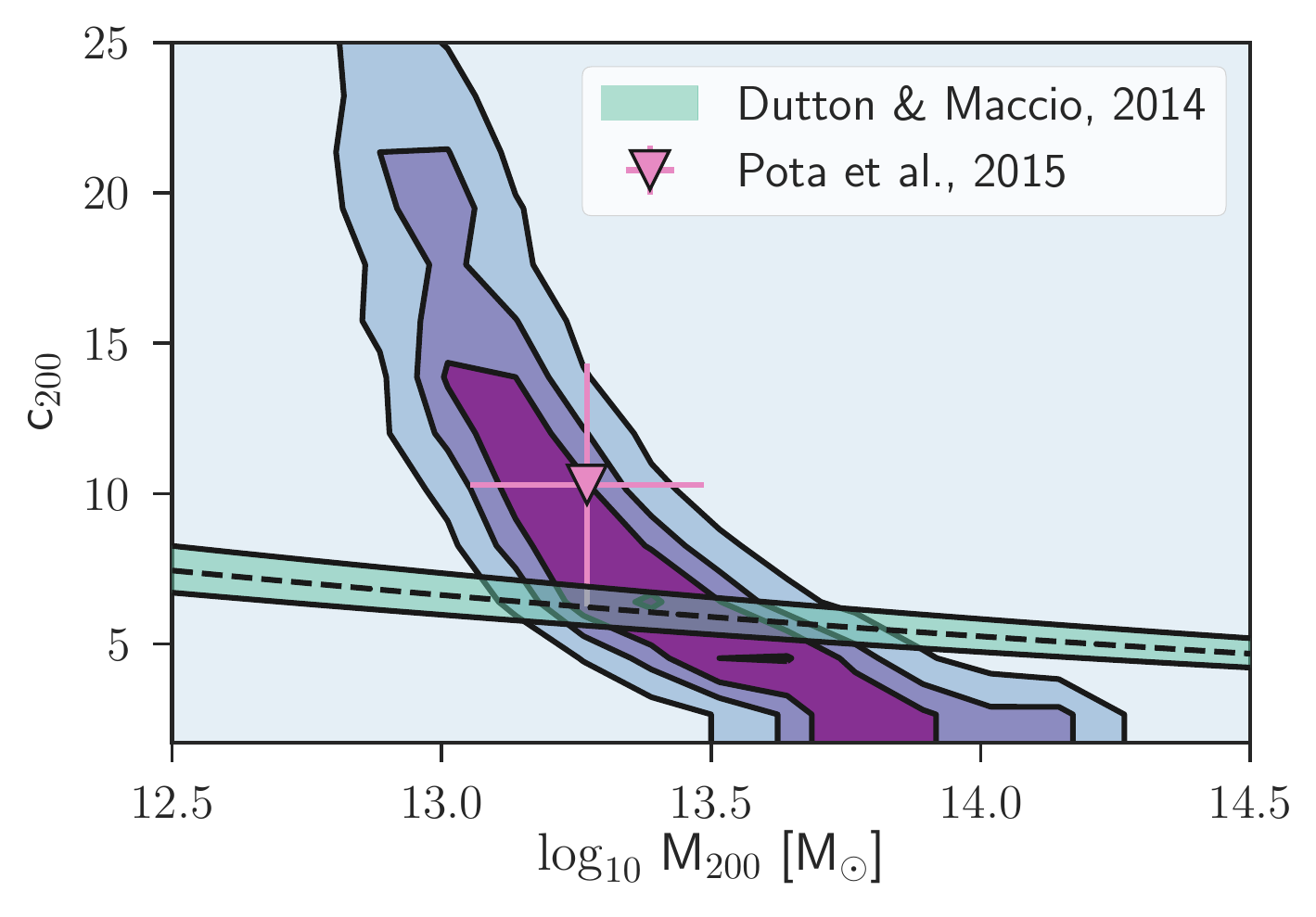}
  \caption{Posterior distribution of halo mass and concentration for NGC~1407 shown in contours.  The green line shows the relation from \cite{dutton&maccio2014} with characteristic scatter.}
  \label{fig:mc_res}
\end{figure}

Given that we find a nearly NFW halo, we compare our viral mass and concentration with the $M_{200}$--$c_{200}$ relation of relaxed NFW halos from \cite{dutton&maccio2014}.  This relation, along with measurements from the literature, are shown in Fig.~\ref{fig:mc_res}.  Here we see good agreement between our median DM halo parameters and those expected from the mass--concentration relation.  \todo{napolitano+2011, ngc 4374}

\begin{figure}
  \centering
  \includegraphics[width=\linewidth]{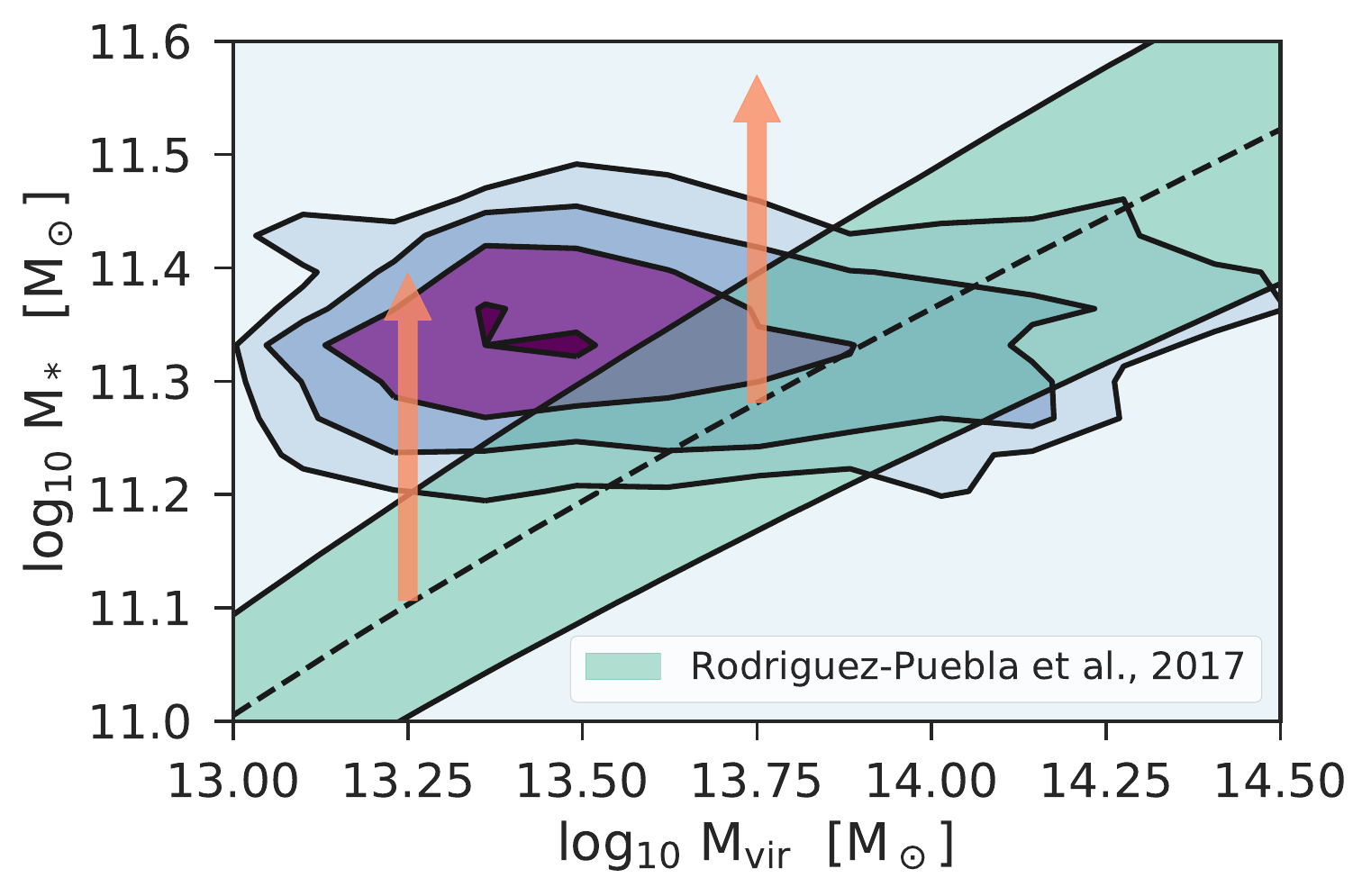}
  \caption{Posterior distribution of stellar mass and halo mass for NGC~1407 shown in contours.  The green line shows the stellar mass-halo mass relation from \cite{rodriguez-puebla+2017} with uncertainties propagated for the 68\% credible interval.  The orange arrows show change in stellar mass from a \cite{chabrier2003} IMF to a \cite{salpeter1955} IMF, with a resulting shift for the prediction of 0.3 dex.}
  \label{fig:smhm_res}
\end{figure}

Fig.~\ref{fig:smhm_res} compares our inference of the halo mass and stellar mass with the $M_*$--$M_\mathrm{halo}$ relation from \cite{rodriguez-puebla+2017}.  Here we have re-calculated our virial mass to match the definition used by \citeauthor{rodriguez-puebla+2017} at $z = 0$, with $h = 0.678$ and $\Delta_\mathrm{vir} = 333$.  We find that NGC~1407 lies slightly above the $M_*$--$M_\mathrm{halo}$ relation from \cite{rodriguez-puebla+2017}.  However, we have indicated the shift in stellar mass that would occur if they had adopted a Salpeter IMF rather than a Chabrier IMF.  We see that this Salpeter IMF stellar mass--halo mass relation is consistent with our inference.

\todo{comparison with shankar+2017, capellari+2015 total mass density profiles}

\subsection{Distance}
One unique aspect of this work is that we freely vary the distance, informed by a weak Gaussian prior from previous redshift-independent distance measurements (see Section~\ref{sec:data}).  With the stellar mass-to-light ratio known to a reasonable degree of uncertainty, this becomes a non-trivial systematic uncertainty, as indicated by the covariance of distance with the inner DM density slope, the stellar anisotropy, and the stellar mass distribution parameters (Fig.~\ref{fig:posterior}).

We find a distance of $21.0^{+1.5}_{-1.4}$ Mpc.  This is a notable offset from our prior distribution on distance, which was a Gaussian with a mean of 26 Mpc and a standard deviation of 2 Mpc.  Our result is inconsistent with the \cite{tully+2013} combined SBF/Fundamental Plane measured distance.  However, our inferred distance is closer to the luminosity distance of $24.2 \pm 1.7$ Mpc at the observed redshift, and in full agreement with the \cite{forbes+2006} distance constraint from modeling the globular cluster luminosity function.  We ran tests fixing the distance to the mean of our prior distribution, and found a lower value of $\gamma$\todo{quote an approximate value here}, consistent with the negative covariance between the two parameters seen in Fig.~\ref{fig:posterior}.  Thus to the extent that our adopted distance is considered low (compared to the wide range of literature values), we find a robust upper bound on $\gamma$.

\subsection{SMBH}

\cite{rusli+2013} modeled the stellar kinematics of 10 ETGs to constrain their super-massive black hole (SMBH) masses.  For NGC~1407, they found $M_\mathrm{BH} = 4.5^{+0.9}_{-0.4}\E{9} \ M_\odot$.  Since we, by design, do not model the detailed dynamics of stellar orbits near the SMBH, we only get weak constraints on its mass.  However, our constraints indicate that the SMBH mass of NGC~1407 could be somewhat lower.  With a uniform prior for $\log_{10} M_\mathrm{bh} < 11$, we find that the posterior distribution on $M_\mathrm{bh}$ cuts off at approximately $3\E{9}$ M$_\odot$.

While \cite{rusli+2013} treated the systematics of having a DM halo in their inference of the SMBH mass, they treated the stellar mass as a constant $\Upsilon_*$ times the stellar luminosity profile.  NGC~1407 lies slightly above standard $M_\mathrm{BH}$--$\sigma$ relation, by a factor of approximately $1.5$ times the intrinsic scatter \citep{mcconnell&ma2013}.  It is conceivable that some of the mass inferred for the SMBH is in fact associated with a more bottom-heavy IMF in the center of the galaxy.

\cite{mcconnell+2013} investigated the effect of radial $\Upsilon_*$ gradients on the inferred masses of SMBHs, finding that a log-slope, $\mathrm{d}\log\Upsilon_* / \mathrm{d}\log r$, which varied from $-0.2$ to $0.2$ had little impact on the inferred $M_\mathrm{bh}$.  However, the radial variation in $\Upsilon_*$ for NGC~1407 appears to be somewhat steeper, with a log-slope of $\sim -0.3$ \citep{vanDokkum+2017}. 

\section{Discussion}
\label{sec:discussion}

\subsection{The $\gamma$--$M_\mathrm{halo}$ relation}
\label{sec:gamma_mhalo}

Few measurements have been made of the inner DM density slope for massive ETGs for reasons discussed in Sec.~\ref{sec:intro}.  Here we discuss both the measurements and predictions for galaxies in this mass regime and for galaxies across a broad range of masses, focusing first on giant elliptical galaxies.

\cite{pota+2015b} also modeled NGC~1407 using GC and stellar kinematics, finding $\gamma \sim 0.6$.  We attribute the difference between this value and our own inference to be primarily due to our more precise determination of the stellar mass distribution and also to fitting distance as a free parameter.  \cite{agnello+2014} modeled the dynamics of the GC system of M87, the Virgo cluster central galaxy.  They found that the behavior of the inner DM density profile followed a power law, $\rho \sim r^{-\gamma}$ with $\gamma \approx 1.6$.  \cite{oldham+2016} also modeled the dynamics of M87, but found evidence for a DM core ($\gamma \lesssim 0.5$).  They attributed this difference to their inclusion of central stellar kinematics in the inference, although we also note that they used a less restricted GC spectroscopic sample than \citeauthor{agnello+2014}.  \cite{zhu+2016} modeled the dynamics of field stars, GCs, and planetary nebulae (PNe) in the massive elliptical galaxy NGC~5846 (based in part on SLUGGS data).  They ran models with a fixed DM core and with a fixed DM cusp, finding a preference for the model with the cored halo.  

\cite{thomas+2007} modeled the stellar dynamics of 17 ETGs in the Coma cluster with both NFW halo models and LOG halo models (which include a central core), though they were unable to distinguish between the two scenarios with the available data.  \cite{napolitano+2010} looked at trends of central DM density and radius for a large sample of low-redshift ETGs, finding evidence for an inner DM density log-slope of $\sim 1.6$, in turn suggesting the need for baryonic processes to contract the halo.  While this result is fairly independent of assumptions about the IMF, it is based on stacked galaxy data and thus it cannot be used to provide $\gamma$ for individual galaxies.

\begin{figure*}
  \centering
  \includegraphics[width=0.9\linewidth]{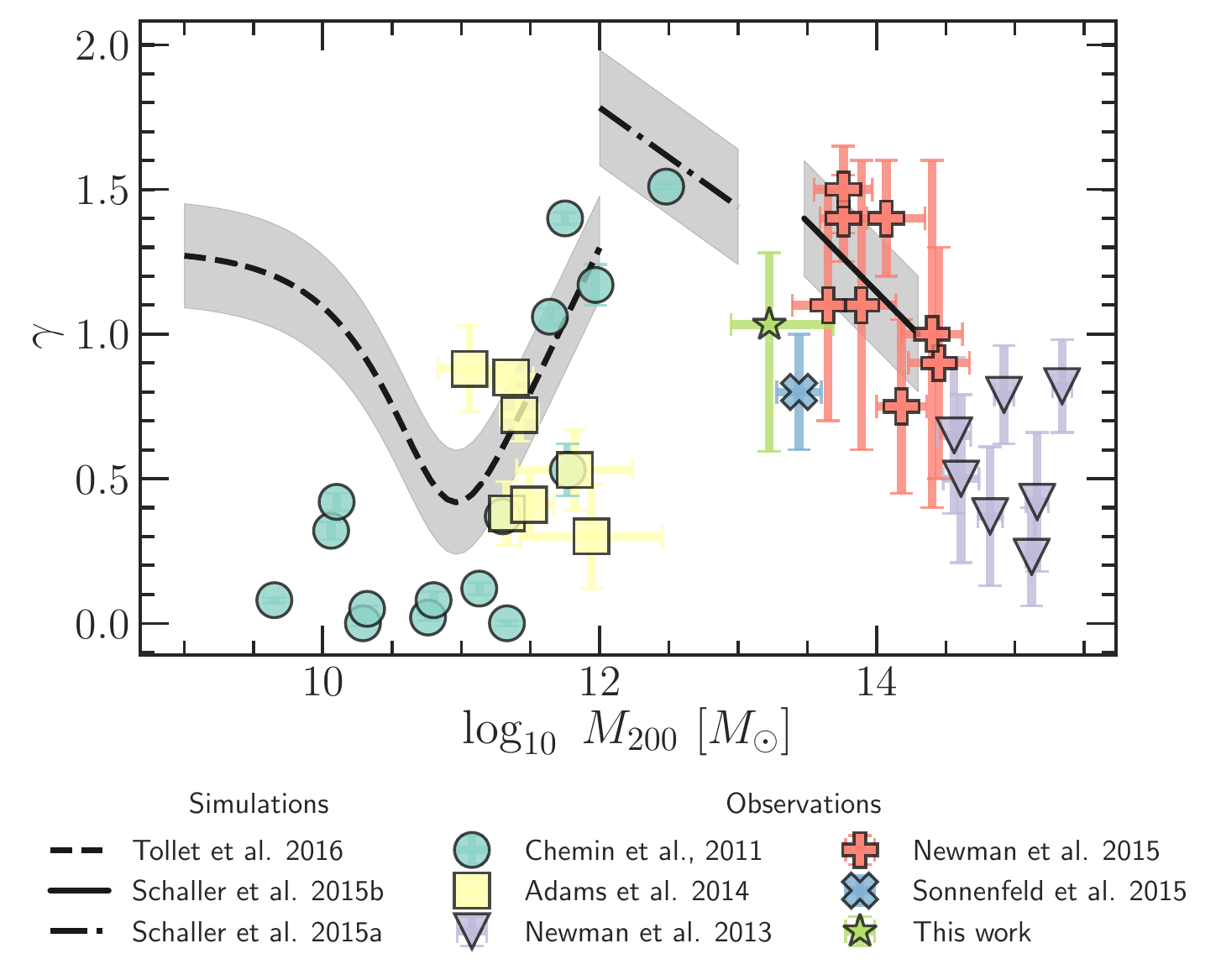}
  \caption{The $\gamma$--$M_\mathrm{halo}$ relation from a wide range of theoretical and observational studies.  We see that $\Lambda$CDM simulations with hydrodynamics (black lines) largely agree with observations (multi-colored points, described in text).  DM core creation occurs most strongly for $10^{11}$ M$_\odot$ halos, with an additional trend towards shallower halos at the highest halo masses.  NGC~1407 follows the general trend of steepening density slope with decreasing halo mass, thought the median value of $\gamma$ is slightly below what would have been interpolated from both theory and observation.}
  \label{fig:gamma_mhalo}
\end{figure*}

In Fig.~\ref{fig:gamma_mhalo} we show how NGC~1407 compares with the observed and predicted dependence of $\gamma$ on halo mass.  We restrict our observational comparisons in this figure to studies which allowed for a variable inner DM density log-slope.  We emphasize that due to the varied definitions, methods of inference, and sources of data used to constrain $\gamma$, Fig.~\ref{fig:gamma_mhalo} is intended merely as a schematic of what we might expect of DM halos across a wide mass range.

We summarize the cited observational studies shown in this figure.  \cite{chemin+2011} modeled the rotation curves of spiral galaxies with Einasto halos.  They reported the log-slope of the best fit Einasto density profile at $\log(r/r_s) = -1.5$, and we compare with their result which assumes a Kroupa IMF.  \cite{adams+2014} modeled the gas and stellar dynamics of dwarf galaxies using both a gNFW profile and a cored Burkert profile.  \cite{newman+2013b, newman+2015} modeled galaxy clusters and groups with constraints from lensing and stellar dynamics with a gNFW profile, finding halos with both with NFW cusps and slightly shallower ($\gamma \sim 0.5$) slopes.

The observations of high mass galaxy clusters suggest a decreasing trend of $\gamma$ with $M_\mathrm{halo}$.  NGC~1407 is consistent with this trend, though it may lie on the turnover region which would be necessary to connect to the increasing trend of $\gamma$ at the low-mass regime.  In subsequent work we will check where this turnover happens with a larger sample of galaxies down to lower halo masses.

Simulations which constrain the relation shown in Fig.~\ref{fig:gamma_mhalo} must address physics across a wide range of spatial scales.  \cite{tollet+2016} used the NIHAO hydrodynamical cosmological zoom-in simulations to make predictions at $\log M_{200} / M_\odot < 12$.  They measured the DM density profiles for their galaxies and reported the log-slope in the region between 1\% and 2\% of the virial radius.  \cite{schaller+2015a, schaller+2015b} used the EAGLE simulations to make predictions for higher mass galaxy clusters. They fitted a gNFW density profile to their halos and reported the inner asymptotic log-slope.  

We see two emerging trends in the $\gamma$--$M_\mathrm{halo}$ relation.  At the range of dwarf and spiral galaxies ($M_\mathrm{200} \sim 10^{10} - 10^{12}$ M$_\odot$), $\gamma$ increases with halo mass.  For hydrodynamic simulations in this regime, DM core creation is associated with bursty star formation \citep{tollet+2016}.  Thus, this trend can be understood as the energy associated with baryonic feedback becoming less and less significant relative to the depth of the potential associated with the halo.  At the range of galaxy groups and clusters ($M_\mathrm{200} \sim 10^{13} - 10^{15}$ M$_\odot$), there is a decreasing trend of $\gamma$ with halo mass.  This has often been interpreted as increased dynamical heating for halos which have experienced more satellite mergers (\citealt{el-zant+2004, laporte&white2015}).

Massive elliptical galaxies like NGC~1407 ought to have the steepest inner density profiles with $\gamma > 1$, owing to the fact that they lie at the intersection of the two competing trends discussed above (i.e., minimal heating from stellar feedback and mergers), and due to the effect of adiabatic contraction (\citealt{blumenthal+1986, gnedin+2004}).

We see that our median value for $\gamma$ falls slightly below the predictions from \cite{schaller+2015a, schaller+2015b} (though consistent within the uncertainty).  However, this value is consistent with the results from the analysis of \cite{sonnenfeld+2015}, who found an average inner DM density slope of $\gamma = 0.80^{+0.18}_{-0.22}$ for a sample of 81 strongly lensed massive ETGs.

The above discrepancy between theory and observation could be an indication that some mechanism is needed to prevent the steepening of the halo density profile. Self-interacting dark matter (SIDM) could be one such mechanism, as the non-zero collisional cross-section allows for heat transfer in the inner regions of the halo.  \cite{rocha+2013} compared the structure of self-interacting DM halos with that of standard CDM halos for two cross-sections, $\sigma / m = 0.1$ cm$^{2}$ g$^{-1}$ and $\sigma / m = 1$ cm$^{2}$ g$^{-1}$.  They found that large cross-sections lead to DM cores within $\sim 50$ kpc.  Our result disfavors this large a cross-section, though we note that it is difficult to rule out their result for the smaller cross-section.  In addition, the lack of baryonic physics in these simulations makes a proper comparison difficult.

\cite{diCintio+2017} used hydrodynamic simulations to explore the effect of SIDM on the baryonic and DM density distributions of Milky Way-mass galaxies.   They used a significantly higher cross-section, $\sigma / m = 10$ cm$^{2}$ g$^{-1}$, than \citeauthor{rocha+2013}.  They reported the log-slope of the density profiles between 1\% and 2\% of the virial radius for both standard CDM simulations and SIDM runs and found a decrease of $0.5-0.7$  in $\gamma$ from the standard run to the SIDM one.

Alternatively, feedback from AGNs could be an important mechanism for transferring energy to the central DM \citep{martizzi+2013, peirani+2017}, analogous to the way that bursty star formation induces potential fluctuations in low mass galaxies \citep{pontzen&governato2012}.  Even in absence of any AGN feedback, dry merging of galaxies can slightly decrease the DM density slope, though not enough to fully counteract the effects of adiabatic contraction \citep{dutton+2015}.

Given the paucity of observational constraints, any connection between our best-fit value of $\gamma$ and any particular physical cause is largely speculation at this point.  However, this ambiguity motivates further work to fill in the remaining observational gaps.

\subsection{Halo anisotropy}
The orbital anisotropy of stars and star clusters in the outer stellar halos of galaxies has received much attention in recent years.  We find that the blue (metal-poor) GCs have tangentially-biased orbits ($\beta_\mathrm{blue} \lesssim -10$), while the red (metal-rich) GCs have radially-biased orbits ($\beta_\mathrm{red} \sim 0.4$).   

Dynamical differences between the red and blue GC subpopulations have been seen before. \cite{pota+2013} calculated the kurtosis of the GC LOSVD as a proxy for orbital anisotropy for a sample of 12 ETGs.  While they found that the kurtosis values for individual galaxies were largely consistent with isotropic orbits, they found that the blue GCs had, on average, negative kurtosis (suggesting tangential anisotropy) in the outskirts while red GCs had, on average, positive kurtosis (suggesting radial anisotropy) in the outer regions.

\cite{pota+2015b} also found tangential blue GCs and radial red GCs for NGC~1407 using Jeans models; we note that we have modeled the same GC dataset as the \citeauthor{pota+2015b} study.

There have been numerous studies of the dynamics of the GC system of M87. \cite{romanowsky&kochanek2001} used the Schwarzschild orbit library method to model the GCs and stars.  They found that the orbits of the GCs as a whole system were near isotropic at large radii.  \cite{zhu+2014} used made-to-measure models to infer the orbits of GCs as a single population, and found orbits that were similarly near-isotropic across most of the spatial extent of the galaxy.  \cite{agnello+2014} found evidence for three GC subpopulations.  For both the bluest and reddest subpopulations they found mildly tangential orbits at 1 $\Reff{}$, while they found the intermediate subpopulation to have slightly radial orbits at the same distance.  \cite{zhang+2015} modeled the dynamics of the red and blue GC subpopulations separately using Jeans models.  They found slightly tangential ($\beta \sim -0.5$) blue GCs in the inner and outer regions of the galaxy, and radially-biased ($\beta \sim 0.5$) red GCs.  \cite{oldham+2016} also modeled blue and red GC subpopulations of M87, finding mildly radially-biased orbits for both blue and red GCs.  Overall the consensus for halo anisotropy in M87 seems to be that, if red and blue GCs have different orbital anisotropies, the blue GC orbits are somewhat more tangentially-biased.


\todo{Zhu+2014 M84 PNe comparison instead of above, comparison with MR GCs in zhu+2016}

\cite{zhu+2016} used made-to-measure models to constrain the $\beta$-profiles of stars, PNe, and GCs in NGC~5846.  They found the opposite trend for this galaxy compared with NGC~1407, with tangentially-biased or isotropic red GCs and radially-biased blue GCs.  The PNe trace the field star population in the center and go from radial to marginally tangential orbits out to $\sim 30$ kpc.

\begin{figure}
  \centering
  \includegraphics[width=\linewidth]{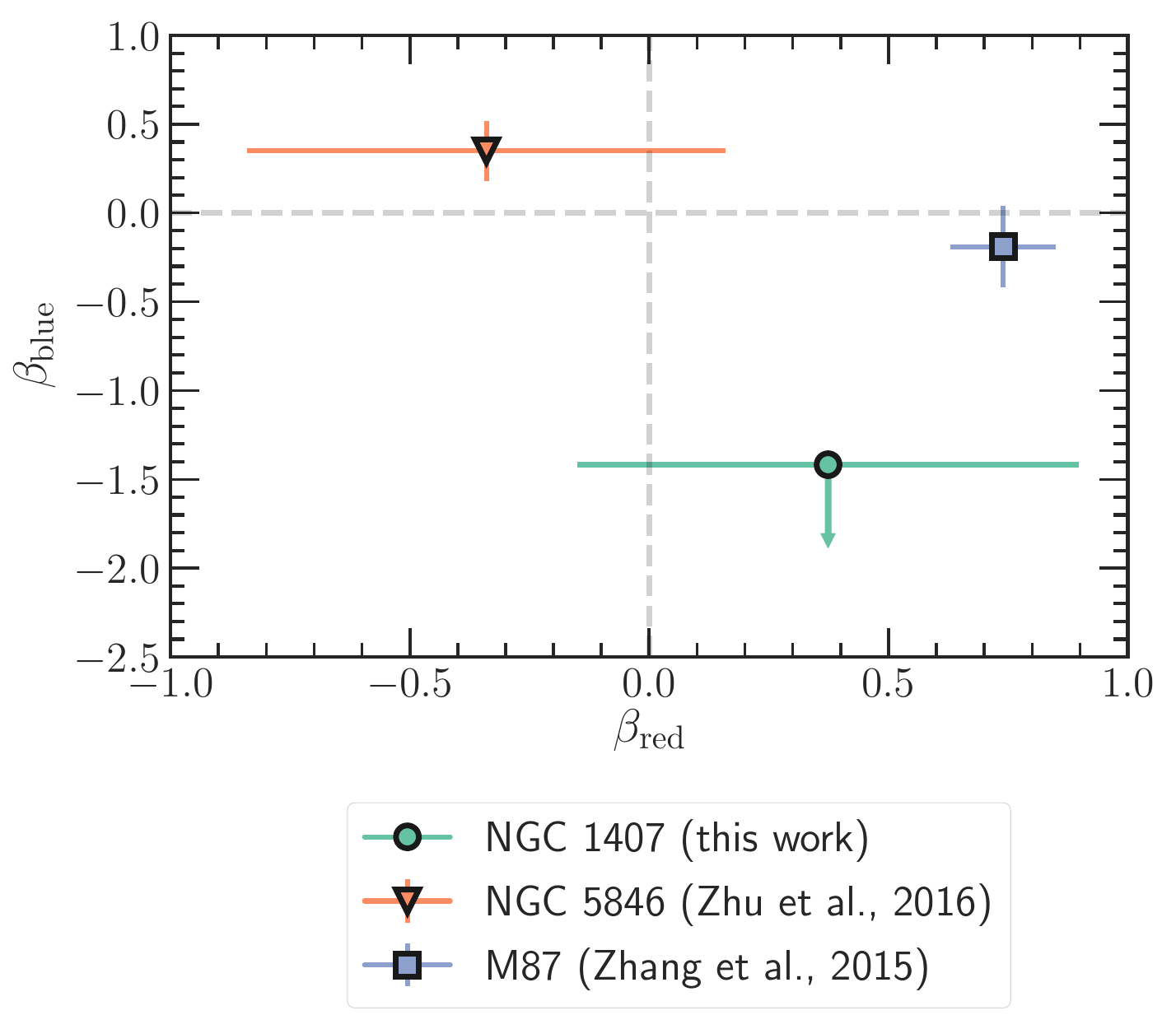}
  \caption{Blue GC orbital anisotropy versus red GC anisotropy for NGC~1407 compared with those of NGC 5846 \citep{zhu+2016} and M87 \citep{zhang+2015}.  For M87, we show their result at a distance of 100 kpc.  Isotropic values of $\beta$ are indicated by the dashed grey line.}
  \label{fig:halo_anisotropy}
\end{figure}

We compare some of these studies which separately analyze blue and red GCs in Fig~\ref{fig:halo_anisotropy}.  There seems to be a diversity of results, with some studies finding the blue GCs to have more tangential orbits than the red GCs, and others finding the opposite result.  However, none of the studies find both red and blue GCs in a single galaxy to have radial orbits (the upper right quadrant of the figure).

This result is puzzling, since the outer stellar halos of galaxies built up by mergers are expected to produce radially-biased orbits (e.g., \citealt{dekel+2005, onorbe+2007, prieto&gnedin2008}), and the majority of blue GCs have most likely been brought into the present day host galaxy via satellite accretions. 

\cite{rottgers+2014} used hydrodynamic zoom simulations from \cite{oser+2010} to examine the connection between orbital anisotropy and the fraction of stars formed in-situ.  They found that accreted stars were more radially biased than in-situ stars.  To the extent that the blue and red GCs could be expected to trace accreted and in-situ populations of the stellar halo, our result that blue GCs have an extreme tangential bias is an interesting counter-example to their result.

One possible explanation for the tangential orbits is that we are seeing a survival-bias effect, whereby GCs on radially-biased orbits are more likely to be disrupted, as they reach more deeply into the center of the potential.  However, for this scenario to work, the metal-poor GCs would have to be in place longer than the metal-rich GCs, contrary to the expectation that the former are accreted and the latter form in-situ.

Another possibility would be a dynamical effect noted by \cite{goodman&binney1984} whereby gradual accretion of mass at the center of a spherical system will preferentially circularize orbits in the outer regions.

The origin of this peculiar halo anisotropy remains an open question, deserving further study.

\subsection{Stellar mass distribution}
\label{sec:discuss_ml}

Since we have chosen to model the stellar mass of the galaxy as its own \sersic{} profile, as opposed to a constant mass-to-light ratio multiplied by the enclosed luminosity, we have a handle on how the stellar mass distribution differs from the stellar light distribution.  We find a half-mass radius of $26^{+4}_{-3}$\arcsec ($2.6^{+0.6}_{-0.4}$ kpc when marginalizing over distance), much smaller than the $B$ band half-light radius of $100$\arcsec ($10.2^{+0.7}_{-0.7}$ kpc).

\begin{figure}
  \centering
  \includegraphics[width=\linewidth]{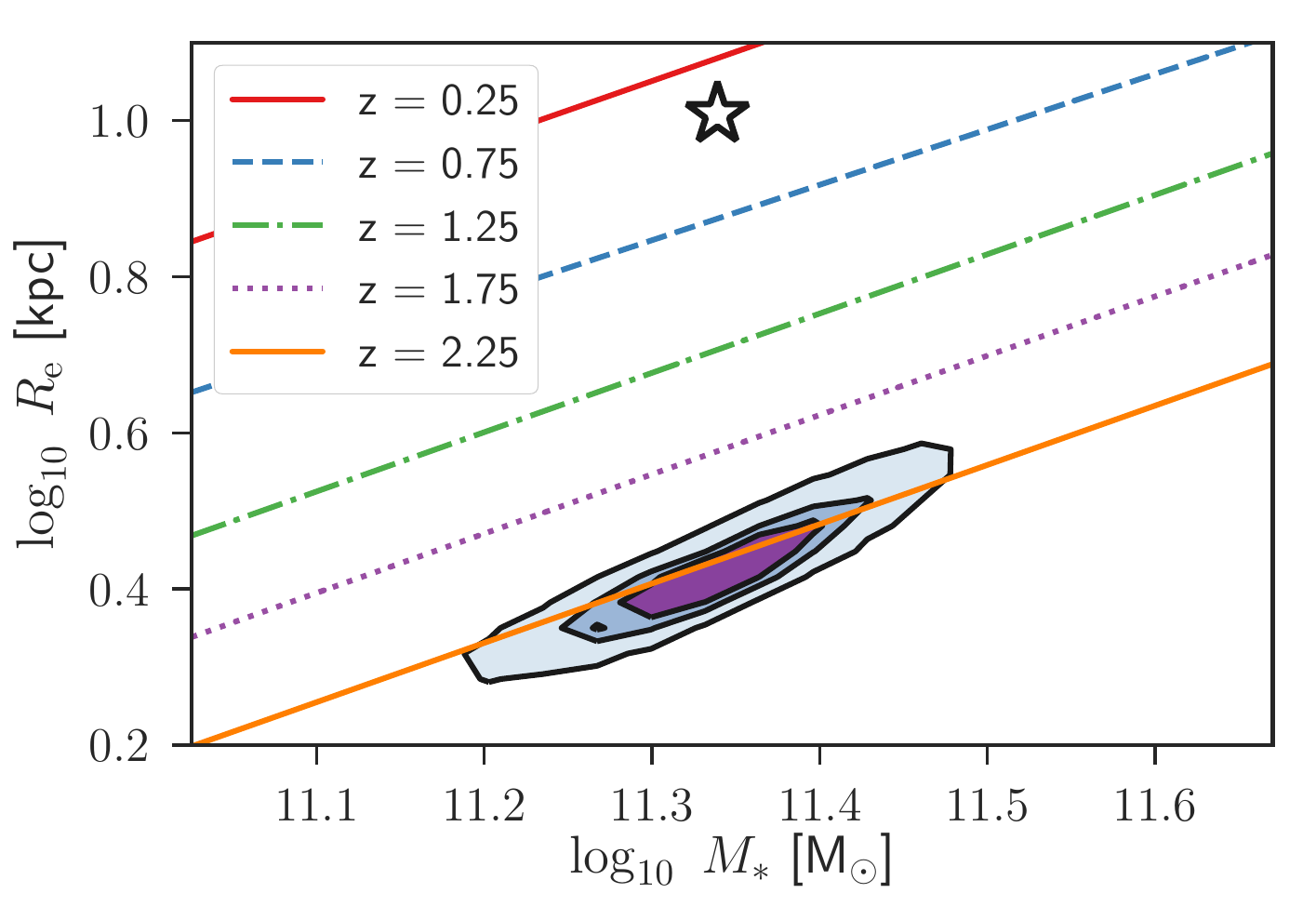}
  \caption{Stellar size--mass relations at different redshifts from \cite{vanderWel+2014}, compared with our inference for NGC~1407 (purple contours).  The $B$ band $\Reff{}$ value of NGC~1407 is indicated by the star.}
  \label{fig:size_mass}
\end{figure}

This relative concentration of the stellar mass is intriguingly similar to the situation at high redshift.  \cite{vanderWel+2014} used results from 3D-HST and CANDELS to trace the evolution in the stellar size--mass relation out to $z \sim 3$, finding a strong size evolution of ETGs at fixed mass of $\Reff{} \propto (1 + z)^{-1.48}$.  We compare our measurement of the stellar half-mass radius with the ETG relations from \cite{vanderWel+2014} in Fig.~\ref{fig:size_mass}.  We see that the stellar mass distribution of NGC~1407 most closely matches the light distribution of compact galaxies at $z \sim 2$.

We note that our modeling may be biased towards smaller effective radii, as we are only fitting to the stellar mass surface density profile where we have data at $R < 100\arcsec$.  If the mass-to-light profile does remain at a Milky Way-like value past $1 \Reff{}$, then this would result in a less compact \sersic{} fit than we find here.

\todo{compare total stellar mass, pota+2015b, forbes+2017b}

\section{Conclusions}
\label{sec:conclusions}

We have presented a new analysis of the dynamics of the massive elliptical galaxy NGC~1407.  We constrained the dynamical mass of the galaxy using a variety of datasets, including metal-rich and metal-poor globular cluster velocity measurements, the stellar velocity dispersion measurements from longslit and multislit observations, and the spatially-resolved mass-to-light ratio from stellar population models.

We found the following:
\begin{enumerate}
\item The dark matter virial mass and concentration are well-matched to expectations from $\Lambda$CDM.
\item The dark matter halo of NGC~1407 likely has a cusp ($\gamma = 1$).  This is shallower than expected for a normal $\Lambda$CDM halo with adiabatic contraction, although a larger sample size is needed to constrain the physical origin of this result.
\item The blue (metal-poor) globular clusters of NGC~1407 are on tangentially-biased orbits (contrary to expectations for accreted stellar mass), while the red (metal-rich) clusters are on slightly radially-biased orbits.
\item The stellar mass distribution is significantly more compact than the stellar luminosity distribution, reminiscent of compact ``red nugget'' galaxies at high redshift.
\end{enumerate}

We are just beginning to probe the $\gamma-M_\mathrm{halo}$ relation in the regime of giant early-type galaxies.  Here we have shown that it is feasible to populate this parameter space with individual galaxies, and we intend to follow up this work with a larger study of galaxies from the SLUGGS survey.

\acknowledgements

This work was supported by NSF grants AST-1211995, AST-1616598, and AST-1616710.  AJR is a Research Corporation for Science Advancement Cottrell Scholar.  AV is supported by the NSF Graduate Research Fellowship Program.  DF thanks the ARC for financial support.  JS acknowledges support from a Packard Fellowship.

Some of the data presented herein were obtained at the W.M. Keck Observatory, which is operated as a scientific partnership among the California Institute of Technology, the University of California and the National Aeronautics and Space Administration. The Observatory was made possible by the generous financial support of the W.M. Keck Foundation.

This research has made use of NASA's Astrophysics Data System and the NASA/IPAC Extragalactic Database (NED) which is operated by the Jet Propulsion Laboratory, California Institute of Technology, under contract with the National Aeronautics and Space Administration.

Funding for the DEEP2/DEIMOS pipelines \citep{cooper+2012, newman+2013_deep2} has been provided by NSF grant AST-0071048.  The DEIMOS spectrograph \citep{faber+2003} was funded by grants from CARA (Keck Observatory) and UCO/Lick Observatory, a NSF Facilities and Infrastructure grant (ARI92-14621), the Center for Particle Astrophysics, and by gifts from Sun Microsystems and the Quantum Corporation.

This research made use of Astropy, a community-developed core Python package for Astronomy \citep{astropy}.  We acknowledge the use of other open source Python packages, including Numpy \citep{numpy}, Scipy \citep{scipy}, Matplotlib \citep{matplotlib}, Pandas \citep{pandas}, IPython \citep{ipython}, and Scikit-learn \citep{scikit-learn}.

The authors wish to recognize and acknowledge the very significant cultural role and reverence that the summit of Mauna Kea has always had within the indigenous Hawaiian community.  We are most fortunate to have the opportunity to conduct observations from this mountain.

\facility{Keck:II (DEIMOS)}

\bibliography{paper}

\pagebreak 

\appendix

\section{Markov Chain Monte Carlo Sampling}
\label{sec:post}

Here we show the detailed results of our sampling of the posterior probability distribution for the model parameter space described in Sec.~\ref{sec:methods}.

\begin{figure}[h]
  \centering
  \includegraphics[width=\linewidth]{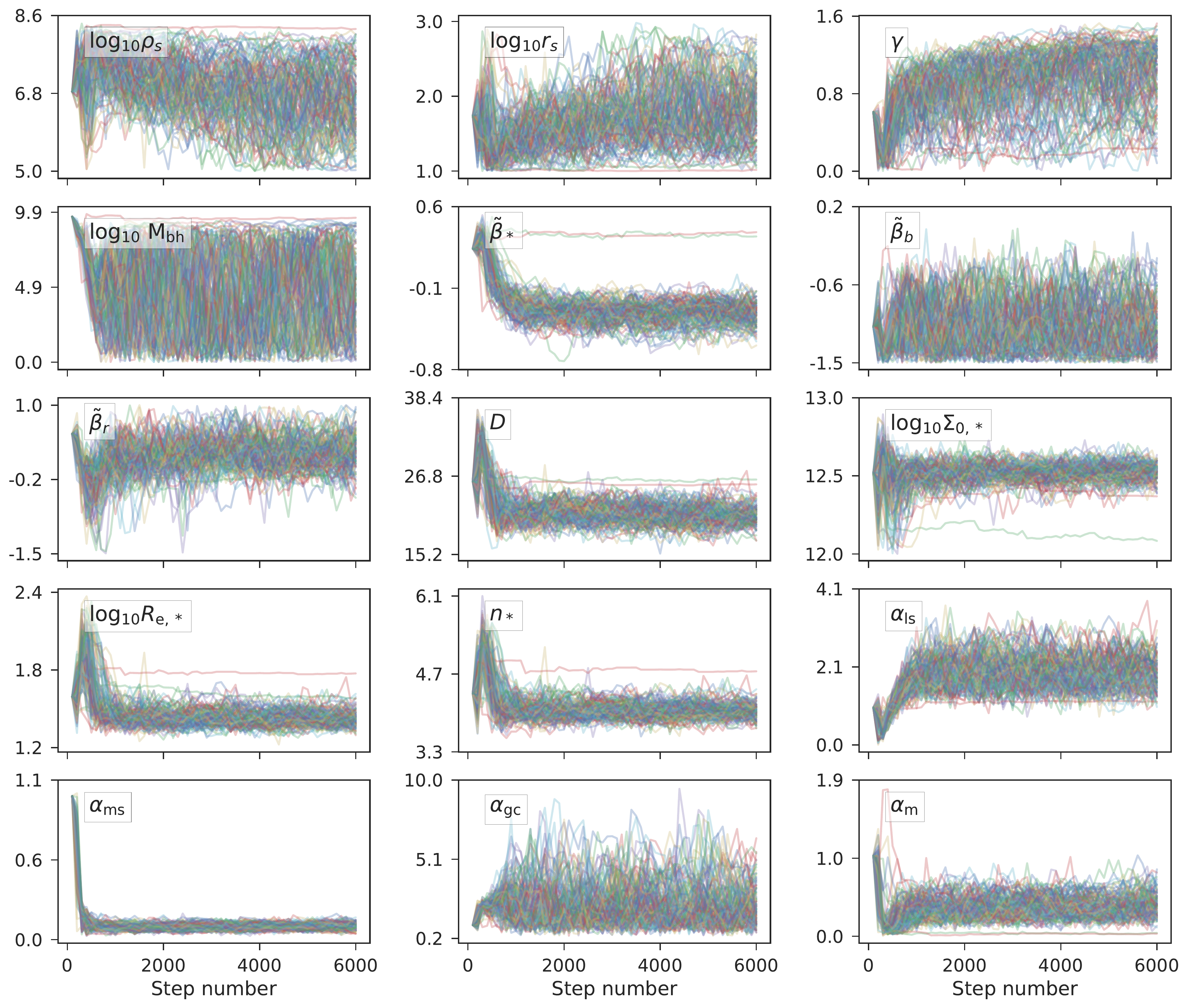}
  \caption{Walker traces across each iteration.  Units are taken from Table~\ref{tab:params}.  $\rho_s$, $r_s$, $M_\mathrm{bh}$, $\Sigma_0,*$, and $R_{\mathrm{e}, *}$ are shown as the logarithm (base 10) of those quantities, and the anisotropy parameters ($\beta$) are shown as the symmetrized anisotropy parameter, $\tilde{\beta} = -\log_{10}(1 - \beta)$.  We reject the first 4500 walker steps in our analysis, where it is clear from the walker traces that the sampler has not yet converged.}
  \label{fig:walkers}
\end{figure}

\begin{figure}[h]
  \centering
  \includegraphics[width=\linewidth]{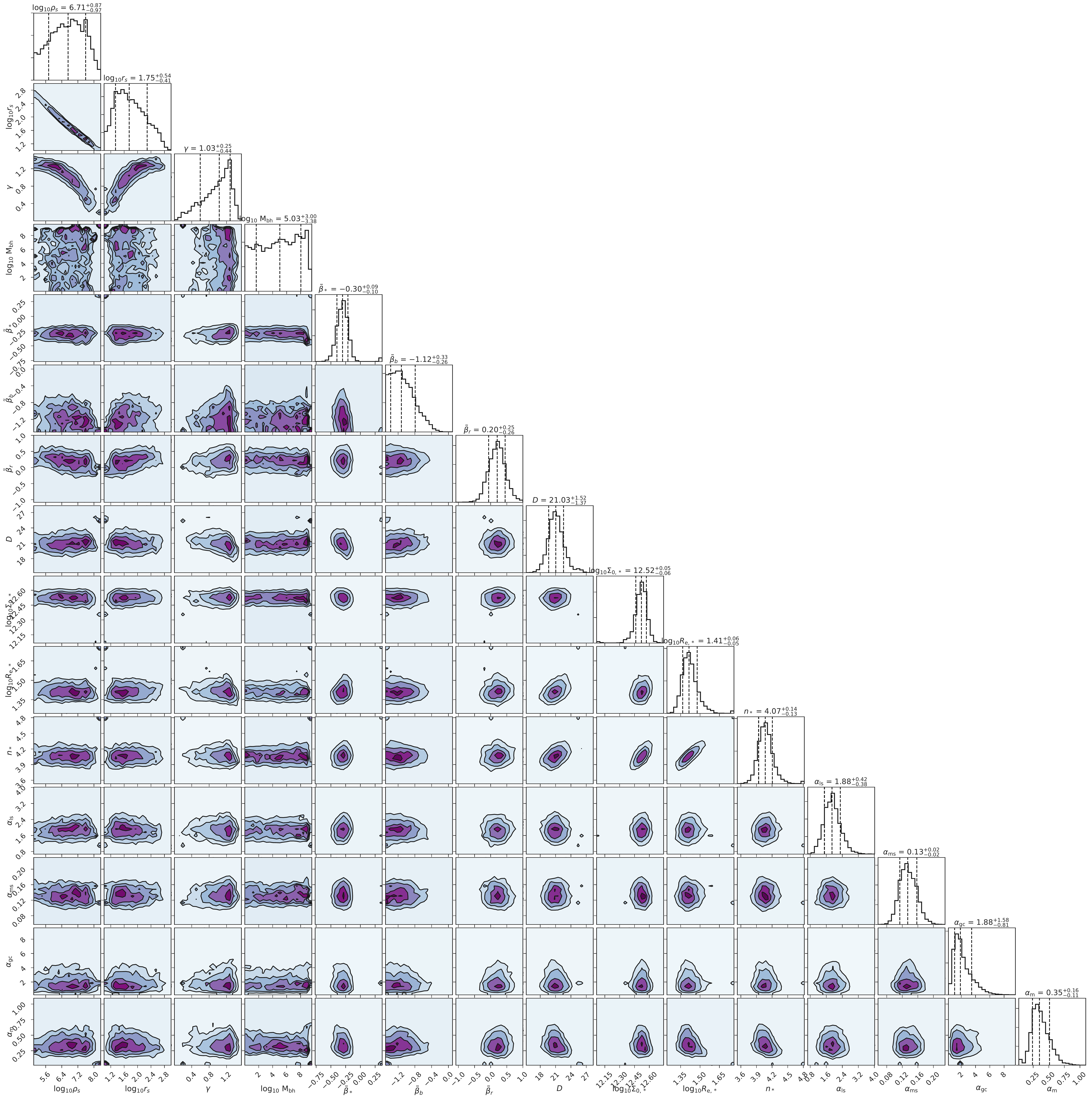}
  \caption{Posterior probability distribution for our model.  Histograms along the diagonal show the marginalized posterior distributions for the respective parameters. The dashed vertical lines mark the 16th, 50th, and 84th percentiles.  The contours (at levels equivalent to 0.5, 1, 1.5 and 2 $\sigma$ for a 2D Gaussian distribution) show the covariances between these parameters.  We hit the prior bounds for $M_\mathrm{bh}$ and $\tilde{\beta}_b$.  For the SMBH, we have very little constraints by design, so we restrict it to be less than $10^{11}$ M$_\odot$.  For all anisotropy parameters, we restrict the range to such that $-1.5 < -\log_{10}(1 - \beta)$ to avoid floating-point underflows.  However, at such tangential orbital anisotropies, the physical differences in the dynamics are negligible.}
  \label{fig:posterior}
\end{figure}

\begin{figure}[h]
  \centering
  \includegraphics[width=\linewidth]{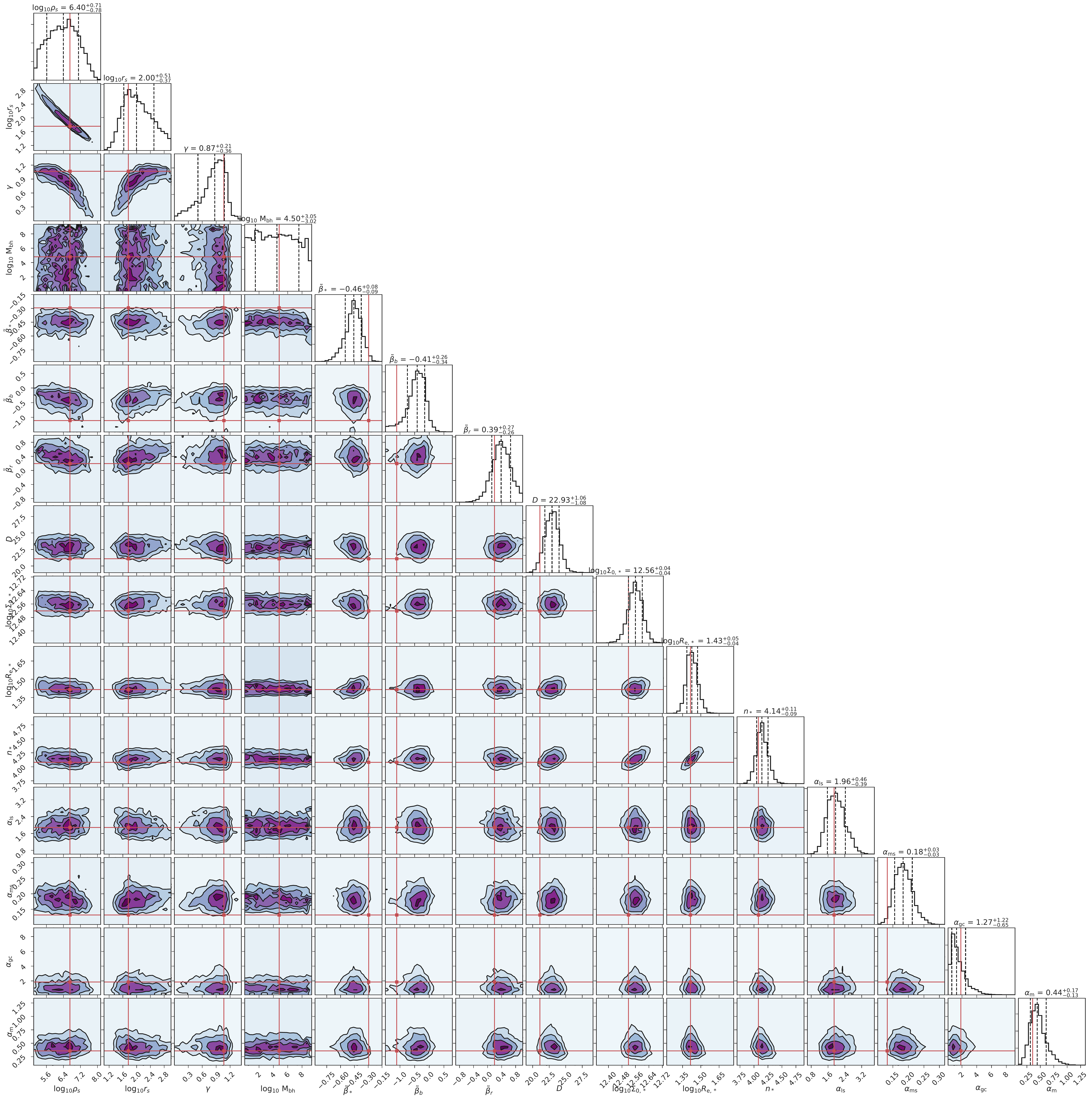}
  \caption{Posterior probability distribution for our model applied to the mock data, as discussed in Sec.~\ref{sec:mock}.}
  \label{fig:mock}
\end{figure}

\end{document}